\documentclass{aa}
\usepackage[varg]{txfonts}
\usepackage[utf8]{inputenc}
\usepackage{amssymb}
\usepackage{lscape}
\usepackage{hyperref}
\usepackage{graphicx}
\usepackage{natbib}
\usepackage{color}
\usepackage{amstext}

\newcommand{\niiab}{[\ion{N}{ii}]$\lambda\lambda$6548,6583}

\newcommand{\siiab}{[\ion{S}{ii}]$\lambda\lambda$6716,6731}
\newcommand{\oii}{[\ion{O}{ii}]}
\newcommand{\oiiab}{[\ion{O}{ii}]$\lambda\lambda$3727,3729}

\newcommand{\oiii}{[\ion{O}{iii}]}
\newcommand{\oiiiab}{[\ion{O}{iii}]$\lambda\lambda$4959,5007}
\newcommand{\oiiia}{[\ion{O}{iii}]$\lambda$5007}

\newcommand{\ciii}{\ion{C}{iii}]}
\newcommand{\ciiiab}{\ion{C}{iii}]$\lambda\lambda$1907,1909}

\newcommand{\mgiiab}{\ion{Mg}{ii}$\lambda\lambda$2796,2803}
\newcommand{\mgii}{\ion{Mg}{ii}}

\newcommand{\haa}{H$\alpha\lambda$6563}
\newcommand{\hb}{H$\beta$}
\newcommand{\hba}{H$\beta\lambda$4861}
\newcommand{\lya}{Ly$\alpha$}
\newcommand{\lyaa}{Ly$\alpha\lambda$1216}
\newcommand{\kms}{km~s$^{-1}$}
\newcommand{\zCOSMOS}{COSMOS-$z_{\rm spec}$}
\usepackage[normalem]{ulem}

\begin{document}

   \title{MAGIC: \textsc{Muse} gAlaxy Groups In \textsc{Cosmos} - A survey to probe the impact of environment on galaxy evolution over the last 8 Gyr
   \thanks{Catalogs described in Appendix \ref{app:catalogs} are available at the CDS via anonymous ftp to \url{cdsarc.u-strasbg.fr} (\url{130.79.128.5}) or via \url{https://cdsarc.cds.unistra.fr/viz-bin/cat/J/A+A/683/A205}}$^{\text{,}}$\thanks{Based on observations made with ESO telescopes at the Paranal Observatory under programs 094.A-0247, 095.A-0118, 096.A-0596, 097.A-0254, 098.A-0017, 099.A-0246, 0100.A-0607, 0101.A-0282, 0102.A-0327, 0103.A-0563.}}
   \titlerunning{The MAGIC survey}

  \author{
    B. Epinat\inst{1, 2}
  \and
    T. Contini\inst{3}
  \and
    W. Mercier\inst{2, 3}
  \and
    L. Ciesla\inst{2}
  \and
    B. C. Lemaux\inst{4, 5}
  \and
    S. D. Johnson\inst{6}
  \and
    J. Richard\inst{7}
  \and
    J. Brinchmann\inst{8, 9}
  \and
    L. A. Boogaard\inst{10, 9}
  \and
    D. Carton\inst{7, 9}
  \and
    L. Michel-Dansac\inst{7}
  \and
    R. Bacon\inst{7}
  \and
    D. Krajnovi\'{c}\inst{11}
  \and
    H. Finley\inst{3}
  \and
    I. Schroetter\inst{3}
  \and
    E. Ventou\inst{3}
  \and
    V. Abril-Melgarejo\inst{12, 2}
  \and
    A. Boselli\inst{2}
  \and
    N. F. Bouché\inst{7}
  \and
    W. Kollatschny\inst{13}
  \and
    K. Kova\v{c}\inst{14}
  \and
    M. Paalvast\inst{9}
  \and
    G. Soucail\inst{3}
  \and
    T. Urrutia\inst{11}
  \and
    P. M. Weilbacher\inst{11}
  }

   \institute{
        Canada-France-Hawaii Telescope, 65-1238 Mamalahoa Highway, Kamuela, HI 96743, USA\\
        \email{epinat@cfht.hawaii.edu}
    \and
        Aix Marseille Univ, CNRS, CNES, LAM, Marseille, France\\
        \email{benoit.epinat@lam.fr}
    \and
        Institut de Recherche en Astrophysique et Planétologie (IRAP), Université de Toulouse, CNRS, UPS, CNES, Toulouse, France\\
        \email{thierry.contini@irap.omp.eu}
    \and
        Gemini Observatory, NSF's NOIRLab, 670 N. A'ohoku Place, Hilo, HI 96720, USA
    \and
        Department of Physics and Astronomy, University of California, Davis, One Shields Avenue, Davis, CA 95616, USA
    \and
        Department of Astronomy, University of Michigan, 1085 S. University, Ann Arbor, MI 48109, USA
    \and
        Univ Lyon, Univ Lyon1, Ens de Lyon, CNRS, Centre de Recherche Astrophysique de    Lyon UMR5574, F-69230, Saint-Genis-Laval, France
    \and
        Instituto de Astrof{\'\i}sica e Ci{\^e}ncias do Espaço, Universidade do Porto, CAUP, Rua das Estrelas, PT4150-762 Porto, Portugal
    \and
        Leiden Observatory, Leiden University, P.O. Box 9513, 2300 RA Leiden, The Netherlands
    \and
        Max-Planck-Institut für Astronomie, Königstuhl 17, 69117 Heidelberg, Germany
    \and
        Leibniz-Institut für Astrophysik Potsdam (AIP), An der Sternwarte 16, D-14482 Potsdam, Germany
    \and
        Space Telescope Science Institute, 3700 San Martin Drive, Baltimore, MD 21218, USA
    \and
        Institut f\"ur Astrophysik and Geophysik, Universit\"at G\"ottingen, Friedrich-Hund Platz 1, D-37077 G\"ottingen, Germany
    \and
        Institute for Astronomy, Department of Physics, ETH Zurich, CH-8093 Zurich, Switzerland
      }

   \date{Received 21 September 2023 / Accepted 1 December 2023}

  \abstract
   {
   Galaxies migrate along filaments of the cosmic web from small groups to clusters, which creates the appearance that the evolution of their properties speeds up as environments get denser.
  }
   {
   We introduce the \textsc{Muse} gAlaxy Groups in \textsc{Cosmos} (MAGIC) survey, which was built to study the impact of environment on galaxy evolution down to low stellar masses over the last 8 Gyr.
   }
   {
   The MAGIC survey consists of 17 Multi-Unit Spectrocopic Exporer (MUSE) fields targeting 14 massive, known structures at intermediate redshift ($0.3<z<0.8$) in the COSMOS area, with a total on-source exposure of 67h. We securely measured the redshifts for 1419 sources and identified 76 galaxy pairs and 67 groups of at least three members using a friends-of-friends algorithm.
   The environment of galaxies is quantified from group properties, as well as from global and local density estimators.
   }
   {
   The MAGIC survey has increased the number of objects with a secure spectroscopic redshift over its footprint by a factor of about 5 compared to previous extensive spectroscopic campaigns on the COSMOS field. Most of the new redshifts have apparent magnitudes in the $z^{++}$ band $z_{\rm app}^{++}>21.5$. The spectroscopic redshift completeness is high: in the redshift range of \oii\ emitters ($0.25 \le z < 1.5$), where most of the groups are found, it globally reaches a maximum of 80\% down to $z_{\rm app}^{++}=25.9$, and locally decreases from $\sim 100$\% to $\sim50$\% in magnitude bins from $z_{\rm app}^{++}=23-24$ to $z_{\rm app}^{++}=25.5$.
   We find that the fraction of quiescent galaxies increases with local density and with the time spent in groups. A morphological dichotomy is also found between bulge-dominated quiescent and disk-dominated star-forming galaxies. As environment gets denser, the peak of the stellar mass distribution shifts towards $M_\star>10^{10}~M_\sun$, and the fraction of galaxies with $M_\star<10^9~M_\sun$ decreases significantly, even for star-forming galaxies.
   We also highlight peculiar features such as close groups, extended nebulae, and a gravitational arc.
   }
   {
   Our results suggest that galaxies are preprocessed in groups of increasing mass before entering rich groups and clusters.
   We publicly release two catalogs containing the properties of galaxies and groups, respectively.
   }

   \keywords{
             Galaxies: high-redshift --
             Galaxies: distances and redshifts --
             Galaxies: evolution --
             Galaxies: groups: general --
             Galaxies: clusters: general --
             catalogs
            }

   \maketitle
%

\section{Introduction}
\label{sec:introduction}

In the paradigm of the $\Lambda$CDM cosmological model, as the Universe evolves, the dark-matter skeleton forms from initial density fluctuations and is composed of voids, walls, filaments,
and nodes. Galaxies are distributed in those large-scale structures, most of which constitute groups and clusters that populate nodes connected by filaments \citep{Bond+96}.
Galaxy environment shapes many galaxy properties, such as their mass, color, star-formation rate, and morphology. Over-dense environments, such as clusters and groups, have a large fraction of elliptical, massive, and red galaxies with a low star-formation activity \citep[e.g.,][]{Dressler80, Hashimoto+98, Weinmann+06}.

Several physical processes have been proposed to explain these changes in galaxy properties seen in local galaxy clusters thanks to mechanisms that act on the gas content and depletion time, such as ram-pressure stripping \citep{Gunn+72}, galaxy--galaxy interactions and mergers \citep{Toomre77}, galaxy--cluster interactions inducing tidal stripping \citep{Byrd+90}, gas starvation or strangulation \citep{Larson+80}, harassment \citep{Moore+96}, or thermal evaporation \citep{Cowie+77}.
These processes may occur once galaxies start their infall onto those structures \citep[as reviewed in][]{Boselli+06}.

According to hierarchical models, the elevated fraction of quenched galaxies in over-dense environments is mainly due to the assembly bias, because galaxies formed earlier in clusters and had more time to interact with their surrounding medium during long-term evolution of proto-clusters into massive clusters.
It has also been suggested that galaxies may be preprocessed before entering the densest structures \citep{Fujita04}, and some observations support this scenario, such as that of \citet{Werner+22}, who report that massive quiescent galaxies at the outskirts of clusters are already quenched at $z\sim 1$.
Numerical simulations seem to indicate that preprocessing is more prevalent in low-mass galaxies \citep{DeLucia+12} and that the halos of former group-like hosts can affect galaxies before their accretion onto a cluster \citep[e.g.,][]{Taranu+14}. Other simulations propose that this process may even start in the walls and filaments of the cosmic web \citep[e.g.,][]{Kuchner+20, Codis+18}, which is consistent with some observations that find some evolution in galaxy properties with respect to the distance to filaments \citep[e.g.,][]{Kraljic+18, Winkel+21, Castignani+22}.

In the low-redshift Universe ($z\lesssim 0.3$), large spectroscopic surveys, such as  the Sloan Digitial Sky Survey \citep[SDSS,][]{Stoughton+02} and the Galaxy And Mass Assembly \citep[GAMA][]{Driver+11}, provide large statistics over wide areas, which provide the means to infer the environment and properties of galaxies down to relatively low mass across a large variety of environments \citep[e.g.,][]{Peng+10}.
At higher redshifts, photometric surveys such as COSMOS \citep{Scoville+07} and CANDELS \citep{Grogin+11}, are relatively efficient thanks to their large-number statistics. Nevertheless, spectroscopic surveys are necessary in order to robustly assess galaxy environments. Large spectroscopic surveys have been undertaken, such as the VVDS \citep{LeFevre+05} and zCOSMOS \citep{Lilly+07, Lilly+09} surveys, from which group catalogs have been produced \citep{Cucciati+10, Knobel+09, Knobel+12}. A large-scale structure at $z\sim 0.7$ with voids, filaments, groups, and clusters was even found in COSMOS \citep[COSMOS Wall,][]{Iovino+16}.
Other spectroscopic surveys have targeted fields around a few massive clusters, such as the GOGREEN and GCLASS surveys used in \cite{Werner+22}, or the ORELSE survey \citep{Lubin+09}, and around groups of  smaller mass, such as the GEEC2 survey with GMOS \citep{Balogh+14}.
All these spectroscopic surveys are nevertheless only representative of the population of galaxies more massive than $M_\star\sim 10^{10}~M_\sun$, owing to their selection functions.

The intermediate redshift range, from $z\sim 0.5$ to $z\sim 1$, is of particular interest for investigating the impact of environment on galaxy evolution, because at $z \gtrsim 0.5$, groups and clusters still contain a large fraction of star-forming galaxies \citep[e.g.,][]{Muzzin+12}, which is due to the long timescale of the processes that turn off star formation \citep[e.g.,][]{Cibinel+13, Wetzel+13}, including preprocessing; it also corresponds to the epoch when groups and clusters start their dynamical relaxation.

To make progress in our understanding of galaxy evolution, it is necessary to elucidate spatially resolved properties,
such as metallicity and kinematics, inferred from spectral features. In particular, the kinematics can be used to relate galaxy integrated and morphological properties to their dynamical mass and angular momentum. Currently, the main samples for which such resolved properties are observed at intermediate to high redshifts are drawn from integral field spectrographs.
Recently, the K-band multi-objet spectrograph (KMOS) has provided samples containing thousands of galaxies (KMOS$^{3D}$, \citealp{Wisnioski+15}; and the KMOS redshift one spectroscopic survey, KROSS, \citealp{Stott+16}).
However, despite their large statistics, those samples are not well suited to study the impact of environment on galaxy properties.
Instead, dedicated surveys are necessary. Some have been achieved, mostly around massive clusters at intermediate redshift with KMOS (\citealp{Bohm+20}; \citealp{Demarco+15}; K-CLASH, \citealp{Tiley+20}; KMOS cluster survey, KCS, and the Virial survey \citealp{Wilman+15, Beifiori+17, Chan+18}), but also with long-slit spectrographs such as VIMOS and DEIMOS \citep{Pelliccia+19}, or FORS2 \citep{Perez-Martinez+17}.
These surveys are either focused on passive galaxies or are limited to relatively massive galaxies due to the necessary preselection of targets, similarly to the spectroscopic surveys mentioned above.
Only wide-field integral field spectrographs alleviate the need for preselection.
This is the case, for instance, for the study of clusters at $z\sim 0.25-0.55$ with SITELLE at
the Canada-France-Hawaii Telescope \citep{Liu+21, Edwards+21} and of a cluster at $z\sim 0.9$ with OSIRIS on the Gran Telescopio Canarias \citep{Perez-Martinez+21}.
In principle, such instruments can also unveil the intracluster or intragroup medium in emission. Nevertheless, these scanning instruments are of limited sensitivity.
Recently, using the Multi-Unit Spectroscopic Explorer \citep[MUSE,][]{Bacon+10}, the Middle Ages Galaxy Properties with Integral field spectroscopy \citep[MAGPI,][]{Foster+21} survey observed 56 fields around massive galaxies at $0.25<z<0.35$ from the GAMA survey in a range of environments. \cite{Richard+21} also observed 12 massive lensing clusters with
MUSE at redshifts $0.2<z<0.6$, some of which host ionized gas nebulae related to ram-pressure-stripping events \citep{Moretti+22}.

While observing clusters of galaxies is of major importance given that they are the most massive virialized structures in the Universe, targeting lower-density environments is also crucial, because galaxy evolution and morphological transformation may start before galaxies enter into clusters, especially in the low-mass regime.
Therefore, the study of the impact of environment on galaxy evolution needs three key elements: (i) environment diversity, given that preprocessing should first occur in walls, filaments, or groups therein; (ii) the low-mass galaxy population, which is the most numerous and may be the population that undergoes the most drastic transformations; and (iii) spatially resolved properties, such as metallicity and kinematics.

Since MUSE has been operating on the Very Large Telescope (VLT), deep fields have proven fruitful for the discovery of groups at intermediate redshifts \citep[$0.3<z<1.5$, e.g.,][]{Bacon+15, Bacon+17, Fossati+19, Leclercq+22, Johnson+22, Cherrey+23}. This is thanks to the sensitivity of this instruments, and to its one square arcminute field of view (FoV), which represents from 250 to 500~kpc from $z\sim0.3$ to $z\sim1.5$. In this redshift range, MUSE is also quite efficient in unambiguously identifying faint and low-mass galaxies \citep[e.g.,][]{Boogaard+18, Bacon+23} and in deriving spatially resolved physical properties from the emission of their ionized gas \citep[e.g.,][]{Contini+16, Bouche+21, Bouche+22}, but also from their stellar kinematics in the most massive cases \citep{Guerou+17}.

Taking advantage of all these unique instrumental capabilities, the \textsc{Muse} gAlaxy Groups In \textsc{Cosmos} (MAGIC), a guaranteed-time-observations large program, was built to investigate how environment has affected galaxy evolution over the last 8 Gyr, by observing 14 massive groups at intermediate redshift ($0.3<z<0.8$). Several studies have already used MAGIC data.
\citet{Abril-Melgarejo+21}, \citet{Mercier+22}, and \citet{Mercier+23} investigated the impact of environment on galaxy scaling relations using spatially resolved kinematics of the ionized gas, which was one of the main goals of the survey.
The MAGIC dataset also serendipitously revealed large ionized gas nebulae \citep{Epinat+18, Boselli+19}, and was used to study the evolution of the galaxy merger fraction since $z\sim 6$ \citep{Ventou+19}; to investigate metallicity gradients of the gas-phase \citep{Carton+18}; to optimally extract spectra of blended sources \citep{Schmidt+19}; and to explore \ion{He}{ii}$\lambda1640$ line properties in high-redshift galaxies \citep{Nanayakkara+19}.

In Sect. \ref{sec:observations} of the present paper, we first present the MAGIC target selection, observing strategy, and data reduction, as well properties inferred from ancillary data. In Sect. \ref{sec:redshifts}, we explain how redshifts were determined and how MAGIC compares to previous spectroscopic campaigns, and assess its spectroscopic redshifts completeness. Section \ref{sec:structures} is devoted to the identification of structures, to the investigation of their properties, and to the description of MAGIC group and galaxy catalogs. In Sect. \ref{sec:properties}, we present an analysis of galaxy population properties as a function of environment, including the red fraction, stellar mass, star formation rate, color, and morphology distributions. Finally, we highlight noticeable features found in MAGIC datacubes in Sect. \ref{sec:peculiar-features}, before summarizing and presenting our conclusions in Sect. \ref{sec:conclusion}.
Throughout the paper, we assume a $\Lambda$CDM cosmology with H$_0=70$ km s$^{-1}$, $\Omega_M=0.7$, and $\Omega_\Lambda=0.3$, and all physical distances mentioned in the paper are proper distances.

\section{MUSE observations, data reduction, and ancillary data}
\label{sec:observations}

\subsection{Target selection}
\label{sec:target_selection}

Observing groups efficiently with MUSE requires to know a priori the position of structures. The goal of our observing program is to observe dense groups at intermediate redshift so that the impact of environment and the size of the sample of galaxies in groups are both maximized.
The knowledge of physical properties of those galaxies such as their mass, their star-formation rate or their morphology is necessary to analyze the impact of environment on galaxy evolution.
We therefore focused our observations on the COSMOS field \citep{Scoville+07}, one of the most explored regions of the sky containing extensive ancillary data and spectroscopic redshifts as complete as possible (see Sect. \ref{sec:comparison_spectro_catalogues}).

Fourteen structures in the redshift range $0.3<z<0.8$ were chosen from the zCOSMOS 20k group catalog \citep{Knobel+09, Knobel+12}\footnote{One additional MUSE field targeting a group at $z\sim 0.706$ from the VIMOS Very Deep Survey (VVDS) group catalog (\citealp{Cucciati+10}, VVDS group number 189) was also initially observed as part of this program. We decide not to include it in the MAGIC sample release for the sake of data homogeneity.}.
We first used this catalog to select groups containing at least five spectroscopic members within one single MUSE field, and further put emphasis on the richest groups including the largest number of emission line galaxies, with at least one emission line with a flux $> 10^{-17}$ erg~s$^{-1}$~cm$^{-2}$. Ten groups were chosen with this strategy, among which, four groups were observed from P94 to P96, with exposures ranging from about 2 to 10 hours.
After these periods, we refined our strategy to prepare for adaptive optics (AO) assisted observations: in P97 to P99, we only observed 1 hour snapshots on fields with suitable tip-tilt stars to further select the most promising candidates for subsequent AO observations.
With this aim, we searched for additional fields, by allowing for mosaicking around large overdensities identified from the zCOSMOS 20k group catalog \citep{Knobel+12} and COSMOS2015 photometric catalog \citep{Laigle+16}.
This added four new fields on two main structures, including a new one (CGr32). After this snapshot campaign, we decided to discard four groups from further AO assisted observations, starting in P100.
We finally decided to add three more groups from the COSMOS Wall group catalog \citep{Iovino+16} for the last P102 and P103 periods. We selected the richest groups that could be observed with AO and that had X-ray counterparts \citep{Gozaliasl+19}\footnote{Two of those groups would have been selected without the X-ray criterion, owing to their richness.}, since the densest groups we previously observed were fulfilling these criteria.

\subsection{Observations}

\begin{table*}
\caption{\label{tab:log}Log of observations for all observed fields.}
\begin{center}
\begin{tabular}{ccccccccc}
\hline\hline
Field      & R. A.           & Dec.        & $\theta$    & Exposure      & Periods     & $\rm FWHM_{PSF}$  & $\partial \rm FWHM_{PSF}/\partial\lambda$ & $\beta_{\rm PSF}$ \\
           & J2000           & J2000       & \degr       & h             &             & \arcsec           & \arcsec / $\mu$m   &  \\
(1)        & (2)             & (3)         & (4)         & (5)           & (6)         & (7)               & (8)   & (9) \\
\hline
CGr23       & 149\degr 47\arcmin 34\arcsec  &  2\degr 10\arcmin 07\arcsec  &     0 & 1             & P98        & 0.615  &  -0.342  &  2.592      \\
CGr26       & 150\degr 29\arcmin 30\arcsec  &  2\degr 04\arcmin 16\arcsec  &    35 & 1             & P98        & 0.586  &  -0.270  &  2.707      \\
CGr28       & 150\degr 13\arcmin 32\arcsec  &  1\degr 48\arcmin 45\arcsec  &   -10 & 4.33 \textbf{(3.33)}   & P98, \textbf{P102}  & 0.569  &  -0.368  &  2.594      \\
CGr30       & 150\degr 08\arcmin 30\arcsec  &  2\degr 03\arcmin 60\arcsec  &    10 & 9.75          & P94, P95   & 0.588  &  -0.282  &  2.606      \\
CGr32-M1    & 149\degr 55\arcmin 14\arcsec  &  2\degr 31\arcmin 53\arcsec  &    30 & 4.33 \textbf{(3.33)}   & P97, \textbf{P100}  & 0.480  &  -0.405  &  2.206      \\
CGr32-M2    & 149\degr 56\arcmin 04\arcsec  &  2\degr 31\arcmin 24\arcsec  &    30 & 4.33 \textbf{(3.33)}   & P97, \textbf{P101}  & 0.490  &  -0.406  &  2.037      \\
CGr32-M3    & 149\degr 55\arcmin 10\arcsec  &  2\degr 30\arcmin 49\arcsec  &    30 & 4.33 \textbf{(3.33)}   & P97, \textbf{P101}  & 0.546  &  -0.498  &  2.241      \\
CGr34       & 149\degr 51\arcmin 24\arcsec  &  2\degr 29\arcmin 26\arcsec  &    -4 & 5.25          & P94, P96   & 0.571  &  -0.236  &  2.825      \\
CGr35       & 150\degr 00\arcmin 21\arcsec  &  2\degr 27\arcmin 23\arcsec  &    30 & 4.67 \textbf{(4.67)}   & \textbf{P102}, \textbf{P103} & 0.555  &  -0.447  &  2.448      \\
CGr51       & 149\degr 58\arcmin 52\arcsec  &  1\degr 47\arcmin 55\arcsec  &   -30 & 1             & P99        & 0.577  &  -0.339  &  2.714      \\
CGr61       & 149\degr 43\arcmin 34\arcsec  &  1\degr 55\arcmin 08\arcsec  &   -40 & 1             & P98        & 0.596  &  -0.344  &  3.320      \\
CGr79       & 149\degr 49\arcmin 07\arcsec  &  1\degr 49\arcmin 19\arcsec  &   -20 & 4.33 \textbf{(3.33)}   & P98, \textbf{P100}  & 0.501  &  -0.345  &  2.474      \\
CGr84-M1    & 150\degr 03\arcmin 03\arcsec  &  2\degr 35\arcmin 48\arcsec  &     0 & 5.25          & P94, P96   & 0.532  &  -0.249  &  2.568      \\
CGr84-M2    & 150\degr 03\arcmin 35\arcsec  &  2\degr 36\arcmin 46\arcsec  &     0 & 4.67 \textbf{(4.17)}   & P97, \textbf{P102}  & 0.608  &  -0.526  &  2.068      \\
CGr87       & 150\degr 01\arcmin 31\arcsec  &  2\degr 21\arcmin 29\arcsec  &   -20 & 2.67 \textbf{(2.67)}   & \textbf{P103}       & 0.540  &  -0.368  &  2.191      \\
CGr114      & 149\degr 59\arcmin 55\arcsec  &  2\degr 15\arcmin 32\arcsec  &   -15 & 4.11 \textbf{(2.17)}   & P94, \textbf{P102}  & 0.588  &  -0.400  &  2.340      \\
CGr172      & 150\degr 10\arcmin 16\arcsec  &  2\degr 31\arcmin 24\arcsec  &     0 & 4.67 \textbf{(4.67)}   & \textbf{P103}       & 0.481  &  -0.483  &  2.074      \\
\hline
\end{tabular}
\tablefoot{(1) Name of the observed field, the field identifier corresponds to the targeted group ID in \citet{Knobel+12}, for CGr32 and CGr84, -M$i$ suffixes correspond to sub-fields of the mosaic. (2), (3) Equatorial J2000 right ascension and declination of the field center. (4) MUSE field orientation measured from North to East. (5) Total on-source exposure time, the value in parentheses with bold fonts corresponds to the exposure time obtained using the adaptive optics mode, if any. (6) ESO periods of observations, bold fonts correspond to AO observations. (7) Image quality at 7000~\AA\ for a Moffat PSF. (8) Image quality variation with wavelength for a Moffat PSF. (9) $\beta$ parameter of the Moffat PSF.}
\end{center}
\end{table*}

\begin{figure*}
 \includegraphics[width=\textwidth]{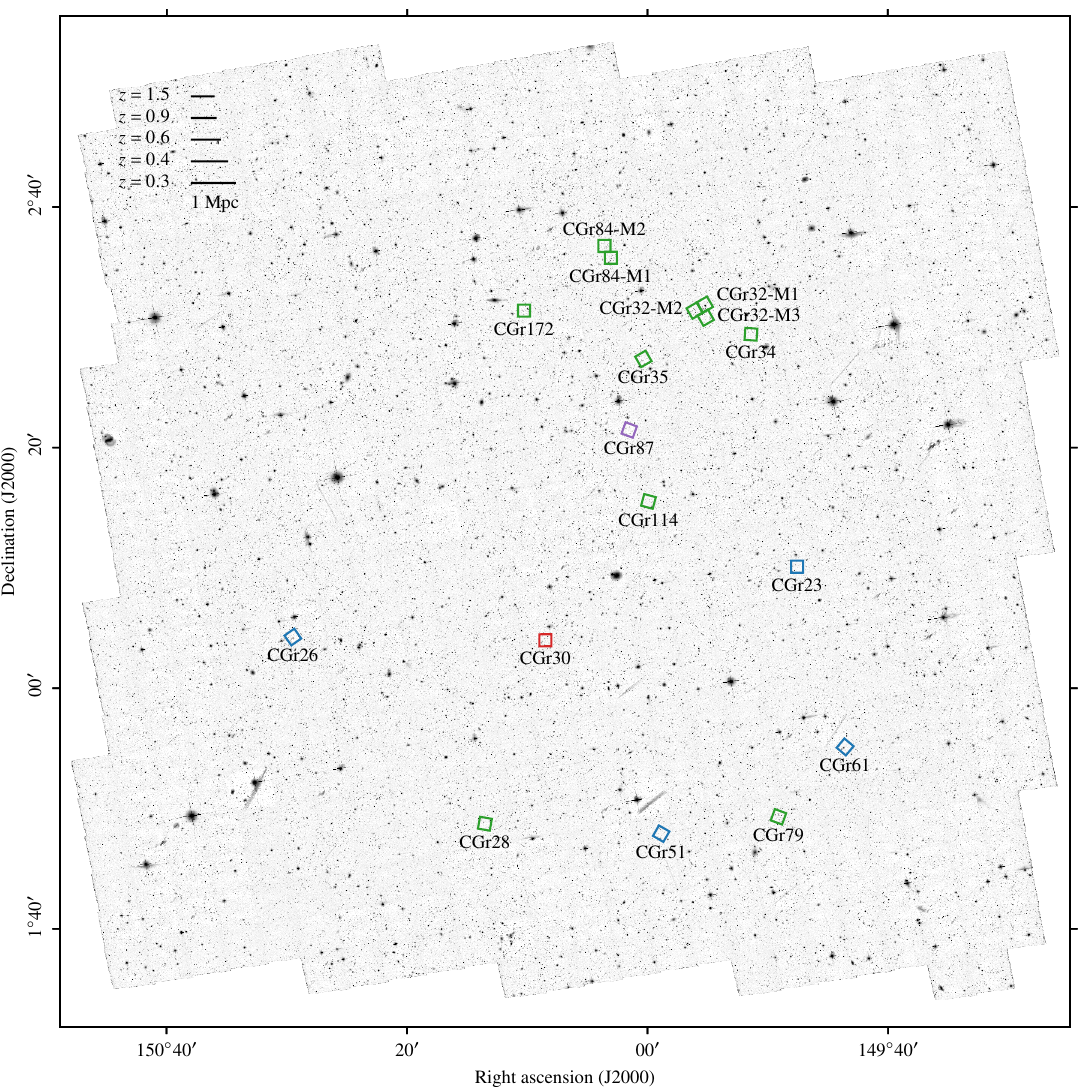}
 \caption{\label{fig:cgr_location}F814W HST-ACS mosaic over the COSMOS field footprint in logarithmic scale (arbitrary unit), with MAGIC fields displayed as squares color-coded according to their depth: blue (1h), purple (2.67h), green (4.11h to 5.25h), and red (9.75h). The scale corresponding to a proper distance of 1 Mpc
 is indicated for various redshifts in the top-left corner.}
\end{figure*}

MUSE observations for the seventeen MAGIC fields were obtained as part of a Guaranteed Time Observations (GTO) program (PI: T. Contini) and were spread over ten observing periods (see Sect. \ref{sec:target_selection} and Table \ref{tab:log}). Their location within the COSMOS field is shown in Fig. \ref{fig:cgr_location}.
For this program, dark nights and good seeing conditions were requested.
For each field, observations were split into observing blocks (OBs) of about 1 hour execution time. OBs consist in four 900s (or 600s) individual exposures with a small spatial dithering pattern ($<0.3$\arcsec) and a 90\degr\ rotation of the field between each individual exposure in order to reduce the systematics induced by the integral field unit (IFU) image slicer. The central coordinates of the fields, their rotation angle with respect to the North, and their total on-source exposure time are summarized in Table \ref{tab:log}. All data were acquired in Wide Field Mode, using either standard seeing-limited or AO assisted observations with the GALACSI AO facility \citep{LaPenna+16, Madec+18}.
Note that with the laser-assisted AO mode, MUSE spectra have a gap in the $5800-5980$\AA\ wavelength range due to the AO notch filter.

Standard calibrations, including day-time instrument calibrations as well as standard star observations, were used for this program. All science exposures include single internal flat-field exposures taken as night calibrations. These short exposures, taken after each OB or whenever there is a sudden temperature change in the instrument, are used for an illumination correction. These calibrations are important to correct for time and temperature dependence on the flat-field calibration between each slitlet throughout the night. In addition, twilight exposures are taken every few days and are used to produce an on-sky illumination correction between the 24 MUSE channels.

The total exposure time devoted to the program reached 91 hours including overheads and 66.7 hours on-source\footnote{95 hours in total and 68.9 hours on-source when including the VVDS field.}. Half the observations (35 hours on source) have used the AO.
Twelve fields have been exposed more than 4 hours and among those, only three did not benefit from AO.

\subsection{Data reduction}
\label{sec:data-reduction}

Each science OB was processed through the MUSE  standard pipeline \citep{Weilbacher+20}. Observations with AO were reduced with the v2.4 version of the pipeline, whereas seeing-limited observations used v1.6, except for the CGr30 field which used v1.2. There are only minor changes between these versions which does not impact the quality of the resulting data cubes. 

Data reduction includes all the common steps to process raw data such as bad pixel tables, bias and dark removals, flat-fielding, wavelength/flux calibration, and sky subtraction. Raw calibration exposures are combined to produce a master bias, master flat and trace table which locates the edges of the slices on the detectors, as well as the wavelength solution for each observing night.
These calibrations are then applied on all the raw science exposures to produce a pixel table without any interpolation. A bad pixel map is used to reject known detector defects, and we make use of the geometry table created once for each observing run to precisely locate the slices from the 24 detectors over the MUSE FoV. Twilight exposures and night-time internal flat calibrations are used for additional illumination correction. Pixel tables are then flux-calibrated and telluric-corrected using standard star exposures taken at the beginning/end of the night. In the case of AO-assisted observations laser-induced Raman lines are also subtracted.
Sky subtraction was applied on each individual exposure. The offsets of each single datacube were computed using stars located in the FoV in order to combine the individual pixel tables into the final datacube.
Sky subtraction was further improved by applying the Zurich Atmosphere Purge (ZAP) software \citep{Soto+16} on the final data cube. ZAP performs a subtraction of remaining sky residuals based on a Principal Component Analysis of the spectra in the background regions of the data cube which has been shown to remove most of the systematics, in particular towards the redder wavelengths. Data reduction produces science-ready data and variance cubes with spatial and spectral sampling of $0.2$\arcsec\ and 1.25~\AA, respectively, in the spectral range $4750-9350$\AA. Astrometry for all the fields was matched to the \textit{Hubble Space Telescope} (HST) Advanced Camera for Surveys (ACS) image coordinates from the COSMOS2015 catalog \citep{Laigle+16}. Average offsets between the MUSE and HST coordinates were computed in the final combined data cube. The data release is explained in Appendix \ref{app:catalogs}.

\begin{figure}
 \includegraphics[width=\columnwidth]{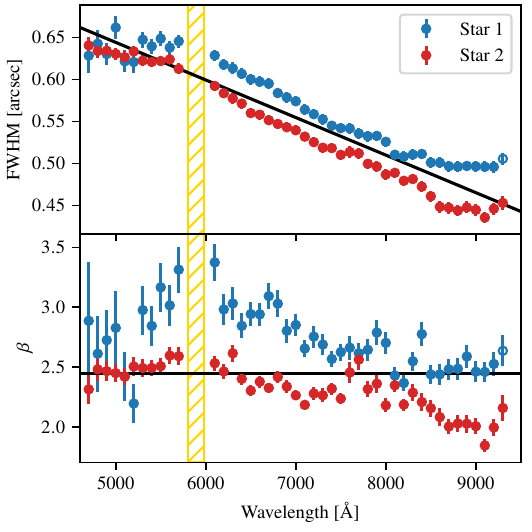}
 \caption{\label{fig:psf_moffat}Moffat PSF fit parameters (FWHM in the top panel, $\beta$ in the bottom panel) as a function of wavelength for the two stars found in the field of CGr35. Open symbols represent outliers discarded from the fit. The linear fit of the FWHM and the mean of $\beta$ are shown as black lines. The absence of points between 5800 and 5980~\AA\ (vertical hatched yellow region) is due to the sodium notch filter when laser-assisted AO observations are acquired.}
\end{figure}

The MUSE line spread function (LSF) is modeled using the results of \citet{Bacon+17} for the MUSE Hubble Ultra Deep Field (MUSE-HUDF) and of \citet{Guerou+17} for the MUSE Hubble Deep Field South:
\begin{equation}
FWHM_{\rm LSF} = \lambda^2\times 5.866\times10^{-8}-\lambda\times 9.187\times10^{-4} + 6.040 \text{~~,}
 \label{eq:LSF}
\end{equation}
where $FWHM_{\rm LSF}$ is the LSF full width at half-maximum, and $\lambda$ is the wavelength, both in \AA.
Following the prescription used for the MUSE-HUDF \citep{Bacon+17, Bacon+23}, we use a Moffat function rather than the Gaussian function used in \citet{Abril-Melgarejo+21} and \citet{Mercier+22} to determine the point spread function (PSF).
We report in Table \ref{tab:log} the PSF parameters that have been computed using objects identified as stars from the spectra in each field. For field CGr87, one bright unresolved object at $z\sim 2.67$ was used in addition to three potential stellar sources having $z=0$ with a poor redshift confidence due to a very low brightness.
We collapse the reduced datacube over 100~\AA\ to create 6\arcsec\ $\times$ 6\arcsec\ images around each star every 100~\AA , and model them using a flat background and a 2D circular Moffat function.
A linear fit is then performed through all measurements after excluding the strongest outliers to determine the variation of the PSF FWHM with wavelength, whereas the Moffat index $\beta$, that does not vary much with wavelength \citep[e.g.,][]{Husser+16}, is determined as the mean across wavelengths of the measurements that converged, weighted by the inverse of their uncertainty. An example is provided for the field of CGr35 in Fig. \ref{fig:psf_moffat}.

\subsection{Galaxy physical parameters from ancillary data}
\label{sec:sed_fitting}

The COSMOS field has been widely observed across a large range of wavelengths, leading to useful ancillary data to infer galaxy properties through photometry and morphology analyses. We hereafter detail how we use these data for the MAGIC sample.

Physical parameters, such as stellar mass, star formation rate
(SFR) and rest-frame magnitudes, are estimated through spectral energy distribution
(SED) modeling using the Code Investigating GALaxy Emission \citep[\textsc{Cigale}\footnote{\url{https://cigale.lam.fr}},][]{Boquien+19}.
\textsc{Cigale} takes into account the energy balance between the ultraviolet (UV) and optical emission absorbed by dust and re-emitted in the infrared and provides a Bayesian-like analysis on the output parameters by building their probability distribution function (PDF).
For each output parameter, the resulting measurement and error are the mean and standard deviation of the PDF.
Single stellar populations (SSPs) of \citep{Bruzual+03} are used with a \citet{Salpeter55} initial mass function (IMF) and a fixed metallicity of 0.02 dex. The star formation history (SFH) is a delayed exponential law with flexibility in the most recent period of the SFH as described in \citet{Ciesla+17} and \citet{Ciesla+21}. This SFH allows for a burst or quenching in the recent SFH with an age and amplitude provided as input parameters. \citet{Ciesla+17} showed that this parametrization yields better stellar mass and SFR measurements than a normal delayed exponential law.
A recent truncation or burst in the SFH is perfectly suited to reproduce the variation of the SFR expected and often observed in galaxies located in high density regions and produced by the interaction with the hostile surrounding environment \citep[][]{Boselli+16}.
The attenuation law is a modified \citep{Calzetti+00} law with a fixed total to selective extinction ratio $R_V=3.1$, leading to five free parameters described in Table \ref{tab:sed_parameters}. The dust template model used is that of \citet{Dale+14}. Nebular and dust templates have default values.

We run \textsc{Cigale} on all galaxies with a tentative or secure redshift (confidence flag larger than or equal to unity, see Sect. \ref{sec:zmeasurements}) in the MAGIC sample. We fix the redshift of each source to that inferred from MUSE data, and take advantage of 32 bands (from UV to radio centrimetric) measured in 3\arcsec\ diameter apertures from the latest COSMOS2020 photometric catalog \citep{Weaver+22}, when available, or from the COSMOS2015 catalog \citep{Laigle+16} otherwise.
We use instantaneous SFR that are provided as output of the SED modeling by \textsc{Cigale}.
Properties derived from this SED modeling and used hereafter are provided in the MAGIC galaxy catalog (see Appendix \ref{app:catalogs}, Table \ref{tab:galaxy_catalog}).

\begin{table}
\caption{\label{tab:sed_parameters}Grid of free parameters for \textsc{Cigale} SED models.}
\begin{center}
\begin{tabular}{lc}
\hline\hline
Parameter &  Values \\
\hline
\multicolumn{2}{c}{\textbf{Delayed exponential + flexibility SFH}}\\
$t_0$ (Gyr)      & 0.001, 0.002, 0.008, 0.023, 0.067, 0.02, 0.6, \\
                 & 1.6, 4.5, 12.9 \\
$\tau$ (Gyr)     & 0.001, 0.003, 0.009, 0.027, 0.081, 0.24, 0.73, \\
                 & 2.2, 6.6, 20 \\
$t_{\rm trunc}$ (Myr) & 1, 12, 23, 34, 45, 56, 67, 78, 89, 100 \\
SFR ratio        & 0.0001, 0.0022, 0.046, 1, 22, 460, 1000 \\
\hline
\multicolumn{2}{c}{\textbf{Attenuation law}}\\
$E(B-V)$         & 0.001, 0.08, 0.16, 0.23, 0.32, 0.39, 0.47, \\
                 & 0.55, 0.62, 0.7 \\
\hline
\end{tabular}
\end{center}
\tablefoot{$t_0$: age of the oldest stars that contribute to the SED; $\tau$: e-folding time of the exponential part of the SFH; $t_{\rm trunc}$: age of the truncation/burst episode; SFR ratio: ratio of SFR after and before the truncation/burst episode; $E(B-V)$: color excess of the nebular lines and of the stellar continuum. Other SED parameters are fixed.}
\end{table}

A morphology analysis was performed in \citet{Mercier+22} for 808 galaxies with secure spectroscopic redshifts in the range $0.25\le z<1.5$ (see Sect. \ref{sec:zmeasurements}). A bulge-disk decomposition was done with \textsc{Galfit} \citep{Peng+02} in order to infer disk inclinations, disk, bulge and galaxy global effective radii, as well as bulge-to-disk ratio (B/D) within the galaxy effective radius. Galaxies for which the fit did not converge were discarded. We refer to \citet{Mercier+22} for the detailed definitions of morphology outputs and to \citet{Mercier+23} for the last updated morphological models.
In addition, updated kinematics parameters of \oii\ emitters obtained from MUSE data using the Moffat PSF are detailed in \citet{Mercier+23}.

Last, the COSMOS field has also been covered in X-ray by \textit{XMM-Newton} and \textit{Chandra} observatories, leading to the detection of groups \citep{Finoguenov+07, George+11, Gozaliasl+19}. The \textit{XMM-Newton} coverage is quite homogeneous over the 14 MAGIC fields from $0.6\times 10^5$~s to $2.3\times10^5$~s, whereas \textit{Chandra} data were mostly used to refine the center of groups in \cite{Gozaliasl+19} thanks to their higher spatial resolution.

\section{Redshifts measurements and completeness}
\label{sec:redshifts}

\subsection{Redshifts determination}
\label{sec:zmeasurements}

\begin{figure}
 \includegraphics[width=\columnwidth]{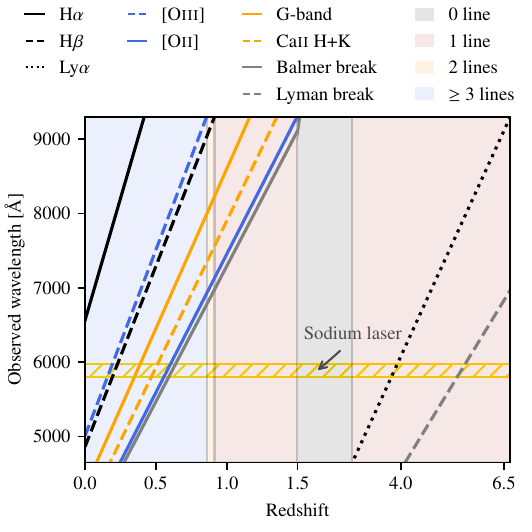}
 \caption{\label{fig:lines_vs_muse_spectralrange}Main spectral features falling in the MUSE spectral range as a function of redshift. Hydrogen Balmer and Lyman lines are shown in black, oxygen lines in blue, absorption lines in orange, and continuum breaks in gray. The background colors indicate the number of bright emission lines simultaneously visible in the MUSE spectral range. As an indication, the spectral window unavailable in AO observations is shown as the horizontal hatched yellow region.}
\end{figure}

For each source in the COSMOS2015 catalog \citep[][]{Laigle+16} within each targeted MUSE FoV, we extracted a PSF-weighted spectrum as described in \cite{Inami+17}.
The redshift estimated from these spectra rely on various spectral signatures such as
(i) emission lines including Balmer series (mainly \haa\ and \hba\ lines, but also higher order Balmer lines when \hb\ leaves the MUSE spectral window above $z \sim 0.9$), the \siiab , \niiab , \oiiiab , \oiiab , \mgiiab\ and \ciiiab\ doublets, and of course the \lyaa\ asymmetric line above $z \sim 2.9$,
(ii) absorption lines such as those related to Balmer decrement, \ion{Na}{i}D$\lambda$5892, \ion{Mg}{i}$\lambda$5175, G-band$\lambda\lambda$4304, \ion{Ca}{ii}~H,K$\lambda\lambda$3934,3968, \mgii , and \ion{Fe}{ii} ($\lambda$2344, $\lambda$2374, $\lambda$2382, $\lambda$2586, $\lambda$2600), and
(iii) continuum features such as D4000, Balmer (3645~\AA), and Lyman (912~\AA) breaks at intermediate and high redshifts, respectively (see Fig. \ref{fig:lines_vs_muse_spectralrange}).

\begin{figure*}
\begin{center}
 \includegraphics[width=0.95\textwidth]{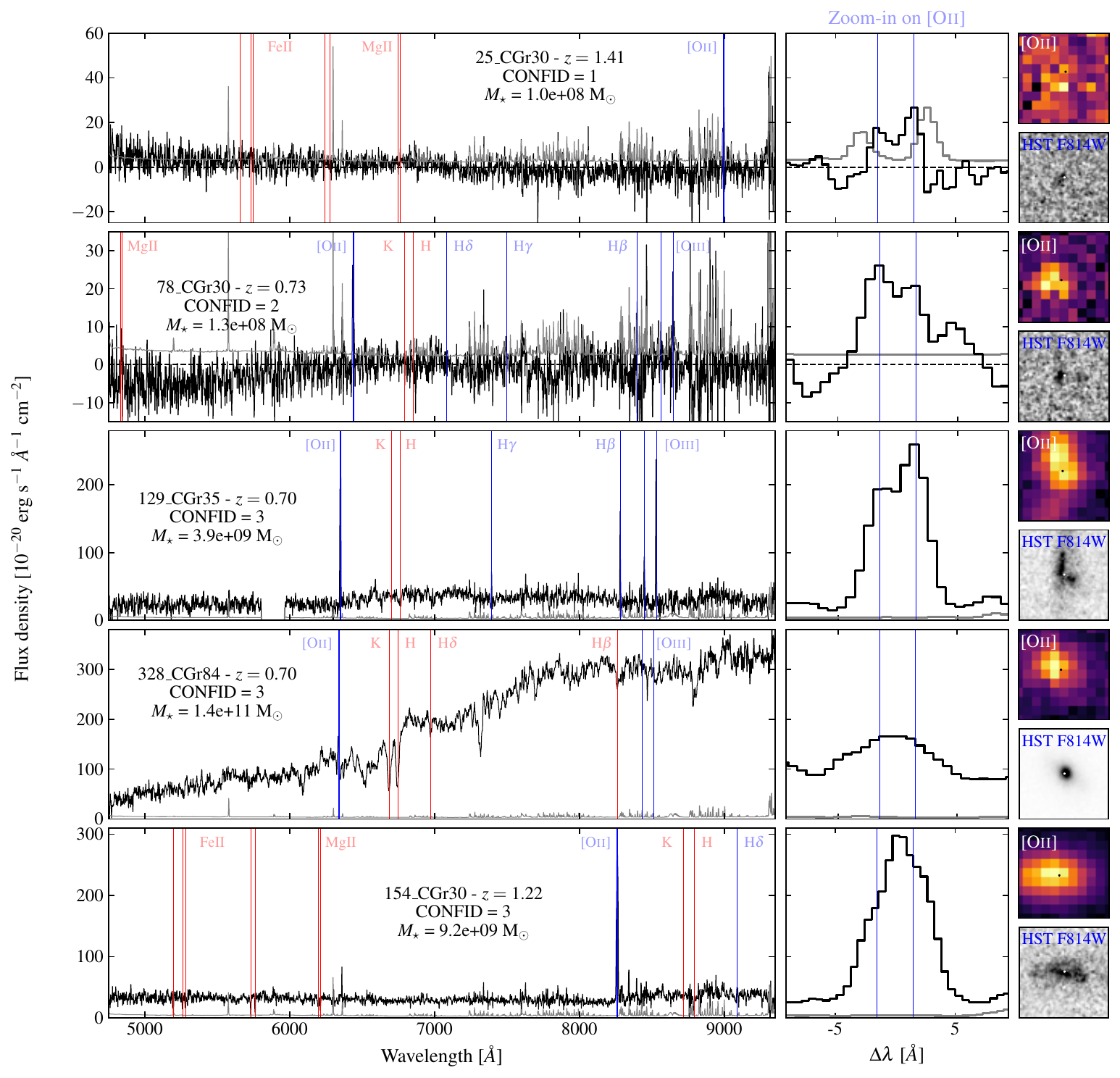}
\end{center}
 \caption{\label{fig:examples_config}Examples of sources with various redshifts and associated confidence levels. From top to bottom, we show spectra (in black) and associated 1$\sigma$ noise (in gray) with (i) confidence 1, with tentatively \oii\ at low
 S/N, (ii) confidence 2 with the \oii\ doublet spectrally resolved with a correct S/N, (iii) confidence 3 with multiple emission lines at high S/N, (iv) confidence 3 with strong continuum features, and (v) confidence 3 with the \oii\ doublet with a high S/N and \ion{Fe}{ii}\ and \mgii\ absorption lines. The main spectral features expected in either absorption or emission are indicated by red and blue vertical lines, respectively. Zoom-in of the normalized rest-frame spectrum around the \oii\ doublet is shown to the right of each spectrum, as well as the MUSE \oii\ narrow band image obtained from collapsing the continuum-free cube over $\pm 5$~\AA\ at rest-frame, and the HST-ACS image in the F814W band. Displayed spectra have been extracted on 1\arcsec\ diameter apertures and smoothed using a two-spectral-element FWHM Gaussian kernel. Images are 2\arcsec\ wide.}
\end{figure*}

Redshifts were estimated with a MUSE customized version of the redshift-finding algorithm MARZ \citep{Hinton+16, Inami+17} using both absorption and emission features.
Redshift precision depends on (i) the nature (emission lines, absorption lines or continuum) and number of spectral features used for redshift determination, (ii) the spectral resolution at the wavelength of those features, (iii) the signal-to-noise ratio (S/N), (iv) the object internal kinematics within the aperture used for redshift determination since it may bias the systemic redshift estimate. Indeed, the absorption lines positions are biased towards the brightest regions, usually at the center of galaxies, whereas emission lines can be dominated by off-center star-forming clumps.
We therefore expect a larger uncertainty for massive galaxies with asymmetric, clumpy and off-centered flux distributions in emission lines, at the exception of those seen nearly face-on. To minimize this effect, we correct the redshift of each spatially resolved galaxy for the systemic redshift derived from the best-fit model of the \oii\ kinematics performed in \citet{Mercier+23}, when available. Given the typical S/N and spectral resolution of MUSE ($R\sim 3000$), the redshift accuracy at 1-$\sigma$ is about $\delta z /(1+z) \sim 0.0001$, corresponding to a 1-$\sigma$ velocity accuracy of $\sim 30$~\kms .

Following other works on deep fields \citep[e.g.,][]{Bacon+15, Bacon+17, Bacon+23}, for each redshift estimate, we assign a confidence level (from 1 to 3) according to its reliability:
\begin{itemize}
 \item Confidence 1: tentative redshift, based on a single emission line or several absorption features with a S/N lower than $\sim 3-5$ per spectral element, either on the emission line maximum intensity or on the continuum.
 \item Confidence 2: secure redshift based on a single emission line (such as the \oii\ doublet) with a S/N above $\sim 5$ without additional information, or several absorption features with intermediate S/N ($\sim 5-10$)
 \item Confidence 3: secure redshift based on multiple high S/N spectral features or on a single recognizable emission line, such as a well resolved \oii\ doublet, or an asymmetric \lya\ line.
\end{itemize}
Examples of spectra with various confidence levels are provided in Fig. \ref{fig:examples_config}.

During redshift determination, we visually inspected integrated spectra, especially those for which MARZ was not able to provide a clear unique solution. This allowed us to identify two sets of spectral features in some spectra due to blended sources. For four spectra, the two sources were both identified in the COSMOS2015 catalog and a further investigation allowed us to determine the redshift of each source accordingly. For fifteen other cases of blending, the secondary source was not in the COSMOS2015 catalog.
We therefore added this source to the MAGIC galaxy catalog either with new coordinates when we were able to disentangle sources in the HST images, or with the exact same coordinates as the main object when this was not possible.

We also searched for objects that were not previously identified in the COSMOS2015 catalog \citep[see e.g.,][]{Epinat+18} in eight MUSE fields (CGr30, CGr32 mosaic, CGr34, CGr84 mosaic, and CGr114). We have ran the ORIGIN emission-line source finding algorithms \citep{Inami+17} on the deepest field (CGr30). For the other fields, we attempted a visual identification of emission lines emitters by scanning throughout the spectral dimension of the continuum subtracted datacubes.
This work was not systematically done on all fields because those additional objects do not have photometric measurements necessary for the main analyses MAGIC was built for.

\begin{figure}
 \includegraphics[width=\columnwidth]{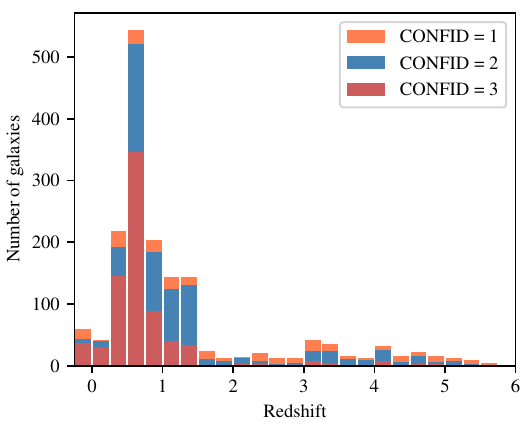}
 \caption{\label{fig:redshift_distribution}Redshift distribution in the MAGIC sample with redshift confidence flags of 1 (orange), 2 (blue), and 3 (red). Stars are shown in the negative bin.}
\end{figure}

\begin{table}
\caption{\label{tab:number_per_class}Statistics for each class of objects as a function of redshift confidence level (CONFID) in the MAGIC survey.}
\begin{small}
\begin{tabular}{lccccc}
\hline\hline
Class &  \multicolumn{5}{c}{CONFID} \\
      & 1 & 2  & 3 & \multicolumn{2}{c}{1-3} \\
\hline
Stars    &     16 (0) &      7 (0) &     37 (0) &     60  (0)&      4\% \\
Nearby   &      2 (0) &     10 (1) &     30 (3) &     42 (4) &      2\% \\
\oii     &    100 (9) &    500 (60) &    654 (46) &   1254 (115) &     75\% \\
Desert   &     43 (4) &     34 (3) &      7 (1) &     84 (8) &      5\% \\
\lya     &    103 (26) &    109 (48) &     31 (10) &    243 (84) &     14\% \\
\hline
All      &    264 (39) &    660 (112) &    759 (60) &   1683 (211) &    100\% \\
\hline
\end{tabular}
\end{small}
\tablefoot{The first four columns indicate the number of objects, whereas the last column indicates the fraction of objects for each class. All sources are included, except those having defects (usually at the edge of MUSE fields) with CONFID equal to unity. Numbers into parentheses correspond to sources found from blind search and deblending. Stars all have $z<0.0005$, nearby galaxies have $0.0005\le z<0.25$, potential \oii\ emitters have $0.25\le z < 1.5$, redshift desert corresponds to $1.5\le z<2.9$ and potential \lya\ emitters have $z\ge 2.9$.}
\end{table}

We ended up with a total of 1683 redshifts with a confidence of at least 1, including 211 galaxies found from blind search or deblending.
The redshifts and their associated confidence levels are given in the MAGIC galaxy catalog (see Appendix \ref{app:catalogs}, Table \ref{tab:galaxy_catalog}).
Figure \ref{fig:redshift_distribution} shows the redshift distribution of galaxies in the MAGIC sample with redshift confidence flags of 1, 2 and 3, whereas Table \ref{tab:number_per_class} summarizes the number of objects in each of the following categories: stars ($z\sim 0$), nearby galaxies ($0.0005\le z<0.25$), \oii\ galaxies ($0.25\le z<1.5$), redshift desert galaxies ($1.5\le z<2.9$) and \lya\ galaxies ($z\ge 2.9$)\footnote{\oii\ and \lya\ galaxies do not necessarily have these lines detected in their spectrum.}. Around 80\% of galaxies with secure redshifts (confidence 2 and 3) are in the \oii\ emitters redshift range. The sharp drop after $z=1.5$ is due to the \oii\ doublet falling beyond the red limit of MUSE spectral range (9350~\AA). Beyond $z=2.9$, there is a rise of secure redshifts due to \lya\ entering the MUSE wavelength range (4750~\AA) with 40\% of those galaxies not being part of the COSMOS2015 catalog. The distribution of redshifts differs from that of the MUSE-HUDF \citep{Bacon+23} for several reasons. First, the MAGIC survey is shallower, since the median (longest) exposure for MAGIC is 4.33 (9.75) hours (see Table \ref{tab:log}), whereas the minimum exposure for the MUSE-HUDF (Mosaic) is 10 hours. Deeper exposures lead to better spectroscopic redshift confidence levels, especially for \oii\ emitters. Secondly, the depth of the COSMOS2015 parent catalog used for source extraction is shallower than that used for the MUSE-HUDF which is based on eXtreme Deep Field (XDF) images \citep{Illingworth+13}. This leads to a limited number of galaxies, especially in the range of \lya\ emitters.
In addition, many galaxies in the MUSE-HUDF were detected blindly and had no corresponding source in the Hubble XDF photometric catalog, whereas for MAGIC, as stated above, such a blind search was not systematically done for all the fields, which also mostly reduces the number of \lya\ emitters.
Last, MAGIC targeted dense groups within the redshift range $0.3<z<0.8$, which explains its quite high fraction of galaxies in this range, especially around $z\sim 0.7$.

\subsection{Comparison with previous spectroscopic campaigns}
\label{sec:comparison_spectro_catalogues}

The COSMOS field has been extensively targeted by several spectroscopic campaigns with VIMOS, such as zCOSMOS \citep{Lilly+07, Lilly+09}, VUDS \citep{LeFevre+15}, COSMOS Wall \citep{Iovino+16}, LEGA-C \citep{vanderWel+16, vanderWel+21}, but also with other spectrographs such as PRIMUS \citep{Coil+11}, DEIMOS (C3R2, \citealp{Masters+19}; 10k, \citealp{Hasinger+18}; \citealp{Kartaltepe+10}; \citealp{Casey+17}), FORS2 \citep{George+11, Comparat+15}, IMACS \citep{Trump+07}, FMOS \citep{Roseboom+12, Kashino+19}, LRIS \citep{Casey+17}, FOCAS \citep{Ikeda+11, Ikeda+12}, HECTOSPEC \citep[hCOSMOS survey,][]{Damjanov+18}, and 3DHST \citep{Brammer+12}, that are part of a new COSMOS spectroscopic redshift master catalog (Khostovan et al. in prep), hereafter referred to as \zCOSMOS\ catalog for simplicity.

We have identified 448 sources having one redshift estimate with a \zCOSMOS\ confidence flag of at least 1 in this master catalog\footnote{A \zCOSMOS\ confidence flag \citep[defined as for zCOSMOS in][]{Lilly+07, Lilly+09} of one refers to an insecure redshift, of two corresponds to a likely redshift with some remaining doubts, and of three or four indicates that the redshift is very secure.} within the footprint of MAGIC fields. Galaxies too close to the edges of MUSE fields to infer a robust redshift were excluded here.
Two galaxies from the \zCOSMOS\ catalog that are not in the COSMOS2015 catalog are missing in the MAGIC catalog\footnote{We added to the MAGIC catalog such other sources for which a redshift could be determined from MUSE data.}.
They both have redshift estimates within the optical redshift desert and are not in any COSMOS photometric survey \citep{Ilbert+09, Laigle+16, Weaver+22}.

Among the 446 remaining sources, 394 (88\%) have at least one likely \zCOSMOS\ redshift estimate (\zCOSMOS\ confidence flag of 2), whereas 428 (96\%) of them have one from MAGIC (confidence defined in Sect. \ref{sec:zmeasurements} larger than or equal to one).
Similarly, 294 (66\%) sources have a secure \zCOSMOS\ redshift estimate (\zCOSMOS\ confidence flag larger than or equal to 3), while 416 (93\%) have such a redshift within MAGIC (confidence of at least 2).
If we restrict our analysis to the 294 objects with secure \zCOSMOS\ redshifts, only twelve do not have a secure spectroscopic redshift from MUSE data, including nine galaxies in the optical redshift desert. The last three galaxies are faint and have no visible emission line in the MUSE datacube matching the \zCOSMOS\ (DEIMOS or VUDS) redshifts.
When both \zCOSMOS\ and MAGIC redshifts are secure, the scaled redshift agreement $\Delta z =(z_{\rm MAGIC} - z_{\rm prev})/(1+z_{\rm MAGIC})$ has a standard deviation $\sigma_{\Delta z} = 0.0005$ after iteratively clipping nine sources at 10$\sigma$. The redshift of one of those sources is that of a blended object that we identified in the MUSE data. We have checked the MUSE spectra of the eight other sources and are confident that we provide the correct redshift solution for all of them.

\begin{figure}
 \includegraphics[width=\columnwidth]{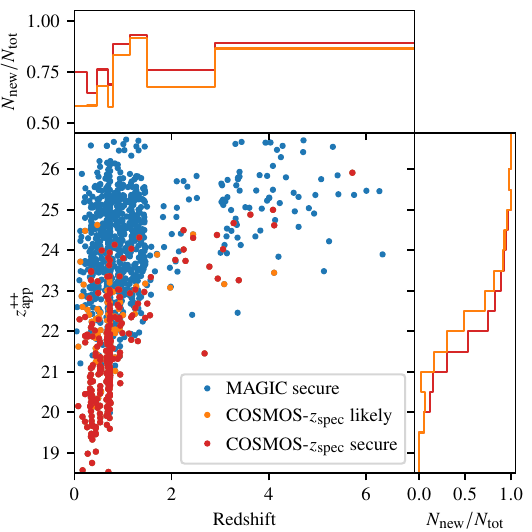}
 \caption{\label{fig:distrib_z_mag}Distribution of galaxies within the MAGIC footprint in the apparent $z^{++}$ band magnitude--redshift plane for \zCOSMOS\ sources with secure (red) and likely (orange) redshifts and new MAGIC sources with secure redshifts (blue).
 The fraction of galaxies with new secure MUSE redshifts is provided as a function of magnitude (right panel) and redshift (top panel), with respect to either \zCOSMOS\ secure (red) or both secure and likely (orange) redshifts.}
\end{figure}

Over its footprint, the MAGIC survey increases the density of objects with a likely redshift by a factor of 4.3 (from 394 to 1683), and with secure redshifts by factor of 4.8 (from 294 to 1419).
To further investigate the strengths of MAGIC in terms of redshift determination, we display in Fig. \ref{fig:distrib_z_mag} each object with new redshifts (blue) and previously secure (red) or likely (orange) redshifts in a redshift-magnitude ($z^{++}$ band) diagram. This figure also shows the corresponding fraction of new MUSE spectroscopic redshifts as a function of magnitude and redshift. Stars are excluded from this figure. Indeed, adding them shows that bright stars were not much targeted in previous spectroscopic campaigns, probably because they are easily identifiable from images.

For bright galaxies, the fraction of new secure redshifts increases from $\sim 0$\% at $z^{++}_{\rm app}\sim 19$ to $\sim 15$\% at $z^{++}_{\rm app}=21.0$ (red curve on the right panel of Fig. \ref{fig:distrib_z_mag}).
For fainter galaxies, the proportion of new spectroscopic redshifts gets larger than 75\% within 1.5 dex and asymptotically reaches 100\%. Most of galaxies with likely \zCOSMOS\ redshifts (orange dots) have $z^{++}_{\rm app}$ magnitudes between 21.5 and 22.5.
The fraction of new secure redshifts reaches a local minimum of about 69\% in the redshift bin $0.69<z<0.79$ because this redshift range was extensively targeted for the study of the COSMOS Wall \citep{Iovino+16}. It may also be that dense environments host galaxies that are forming less stars, with fainter emission lines, and therefore more difficult to attribute a secure spectroscopic redshifts in the low-mass regime (see Sect. \ref{sec:galaxies_properties}).
Apart from this redshift range, the fraction of new galaxies increases linearly with redshift from about $65$\% at $z=0.25$ to nearly $93$\% at $z=1.5$.
This is mainly due to the fact that the fraction of bright galaxies decreases with redshift.
The higher spectral resolution of MUSE could also contribute to this trend by better identifying the \oii\ doublet when it is the only available feature, and by slightly reducing night sky lines contamination in the red part of the spectrum.
The fraction of new redshifts in the redshift desert is above 75\%, and around 90\% beyond $z=2.9$. Running automatic source finding algorithms such as ORIGIN would further increase the number of new redshifts in these redshift ranges.
These findings highlight that, owing to their selection functions, the sampling rate of previous spectroscopic campaigns was quite high for galaxies brighter than $z^{++}_{\rm app}=22.5$, and demonstrate the strength of wide spectral range integral field spectroscopy to determine robust redshifts for faint sources without any magnitude-limited selection.

\subsection{Spectroscopic completeness}
\label{sec:completeness}

To further assess how the MAGIC sample is representative of the population of galaxies at various redshifts, we analyze the spectroscopic completeness of the survey by measuring the fraction of galaxies with a secure spectroscopic redshift (confidence 2 and 3) with respect to the actual number of galaxies within given redshift and magnitude bins.
We estimate the latter thanks to the COSMOS2015 photometric catalog \citep{Laigle+16} that uses $z^{++}$ band images with a magnitude limit of 25.9 in 3\arcsec\ diameter apertures for source detection, ensuring a completeness close to unity up to this limit.

\begin{figure}
 \includegraphics[width=\columnwidth]{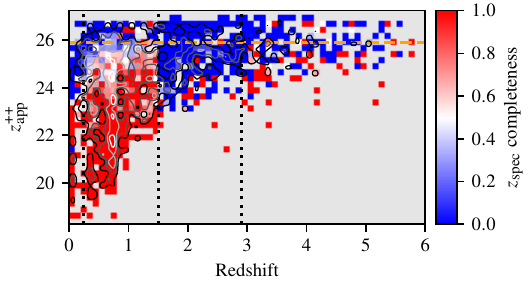}
 \caption{\label{fig:completeness}Spectroscopic redshift completeness as a function of redshift and $z^{++}$ band apparent AB magnitude measured in 3\arcsec\ diameter circular apertures.
 The black vertical dotted lines show the redshifts $z=0.25$, 1.5, and 2.9, whereas the orange dashed line corresponds to the limiting magnitude $z^{++}= 25.9$ of COSMOS2015 measured within 3\arcsec .
 Contours correspond to the number of galaxies per bin with photometric redshifts at levels 2, 4, 8, and 16 (from black to white).
 Beyond $z=1.5$, only a few discrete completeness values are computed due to the limited statistics within each bin.}
\end{figure}

Figure \ref{fig:completeness} shows the local spectroscopic redshift completeness as a function of both photometric redshift and $z^{++}$ band apparent magnitude, in bins of width $\Delta z = 0.1$ and $\Delta z^{++} = 0.25$, respectively.
Before computing the completeness, we first replace photometric redshifts by MAGIC spectroscopic redshifts, when available and secure. Indeed, as underlined in \cite{Laigle+16}, the photometric redshift error inferred from $\Delta z= (z - z_{\rm phot})/(1 + z)$, with $z$ and $z_{\rm phot}$ being the actual and photometric redshifts, respectively, increases when galaxies become fainter and gets as large as 0.05 at 1$\sigma$ for the faintest magnitudes ($i_{AB}>25$). Doing such a correction is particularly important to avoid smoothing out artificially completeness around redshift bins where spectral signatures leave or enter the MUSE spectral range (see Fig. \ref{fig:lines_vs_muse_spectralrange}), especially at $z\sim 1.5$ and $z\sim 2.9$.
There might nevertheless remain boundary effects around those redshifts.
Indeed, while the above correction adds galaxies with a photometric redshift $z>1.5$ ($z<2.9$) to the bin below $z=1.5$ (above $z=2.9$), it does not remove from this bin galaxies that actually have $z> 1.5$ ($z<2.9$) but do not have strong spectral signatures to infer a secure spectroscopic redshift. Therefore, the parent population in the bin below $z=1.5$ (above $z=2.9$) may be slightly overestimated and the completeness underestimated.
Those effects are nevertheless negligible compared to not performing the correction, especially where completeness is high.

The main result is that completeness is globally close to 80\% below $z=1.5$, considering all galaxies with $z^{++}_{\rm app}\le 25.9$, and drops sharply down to around $10$\% beyond this redshift. This is due to the \oii\ doublet leaving MUSE spectral range, and to the \ciii\ doublet being fainter and more difficult to unambiguously identify. Beyond $z\sim 2.9$, completeness increases to roughly $40$\%, due to \lya\ entering the MUSE spectral domain.
The redshift bin $0\le z <0.1$ contains 60 and 45 sources with photometric and secure spectroscopic redshifts respectively, 41 among the latter being stars. Spectral signatures of stars are identified unambiguously down to $z^{++}_{\rm app}=23$ as shown in Fig. \ref{fig:completeness} by the transition from a null to 100\% completeness around that magnitude.
More generally, completeness is close to unity for objects with an apparent magnitude lower than 23, whatever the redshift, since such sources have either (i) strong emission lines or (ii) detectable signatures in the continuum for the brightest ones. Indeed, the continuum spectra of objects brighter than $z^{++}_{\rm app}=22$ have a $\rm S/N>5$ per angstrom for a typical depth of 5h.
For galaxies with redshift $0.25\le z<1.5$, the completeness remains locally higher than 50\% down to an apparent magnitude of roughly $25.5$ since most galaxies are star-forming in this regime, though SFR decreases at low mass, leading to the detection of potentially strong emission lines within the MUSE wavelength range (see Sect. \ref{sec:zmeasurements}).
The completeness at low apparent magnitudes is slightly increasing with redshift below $z=1.5$.
This is due to the fact that galaxies probed in this magnitude regime have increasing stellar masses, hence increasing SFR and ionized gas line fluxes with redshift. This is probably stepped up by the increase in cosmic SFR with redshift \citep[][]{Hopkins+06}.
Spectroscopic completeness is also slightly lower at low mass at $z\sim 0.6 - 0.8$.
This may be related to the MAGIC observing strategy, targeting many dense structures in this redshift range. Galaxies in these massive groups are expected to have a reduced star-formation rate at a given mass, hence reducing the strength of emission lines below the detection limit for the least massive ones, as shown in \cite{Mercier+22}.

We are further able to estimate spectroscopic redshift completeness as a function of stellar mass, based on masses inferred from SED fitting, either from the COSMOS2015 catalog when no redshift can be derived from the MUSE data, or from our \textsc{Cigale} estimates otherwise.
One drawback for our analysis is that \cite{Laigle+16} only provide the whole stellar mass completeness of the photometric sample, and not in bins of stellar mass. They showed that the 90\% completeness is obtained for stellar masses increasing from $\sim 10^{8.5}~M_\sun$ to $\sim 10^{9.5}~M_\sun$ in the redshift range between $z=0$ and $z=1.5$. The parent photometric sample therefore lacks secure constraints down to low masses reached within MAGIC.
Nevertheless, using a cut in $z^{++}_{\rm app}<25.9$, we can state that the parent sample has a completeness close to unity from $z=0.25$ to $z=1.5$ at least down to a stellar mass limit increasing from $10^{8}~M_\sun$ to $\sim 10^{9}~M_\sun$. Within bins of stellar mass as low as these values, the spectroscopic redshift completeness in the whole redshift range is larger than 60\% and 70\% respectively, and is up to 90\% for galaxies within bins of stellar mass above $10^{10}~M_\sun$.

The effect of depth on completeness was also investigated by analyzing the 1 hour exposure fields only (cf. Table \ref{tab:log}). Despite lower statistics, results are quite similar to those obtained for the full sample, at least within the $z<1.5$ range. The difference in magnitude where a spectroscopic completeness of 50\% is reached is of the order 0.25 dex. The main consequence of exposing more than 4h is to increase completeness for objects fainter the COSMOS2015 limiting magnitude $z{++}=25.9$, and in the redshift range between 1.5 and 2.9.

The results we obtain for MAGIC are comparable to those obtained by \cite{Bacon+23} in the MUSE-HUDF on the Mosaic (10h exposure). We have computed the completeness in the MUSE-HUDF Mosaic using only galaxies brighter than an apparent magnitudes $\text{F775W}=25.8$, similar to the limit of the COSMOS2015 catalog. The main differences are that the completeness in MAGIC is higher than in the MUSE-HUDF in the redshift range of \oii\ emitters ($\sim 75$\% vs $\sim 55$\%), and that it is lower in the redshift desert ($\sim 15$\% vs $\sim 35$\%) and for \lya\ emitters ($\sim 40$\% vs $\sim 70$\%). The latter is mainly due to the overall lower depth of MAGIC data associated to the systematic use of blind searching algorithms for the MUSE-HUDF, since most galaxies above $z=1.5$ have low magnitudes. The higher completeness in MAGIC for \oii\ emitters is related to the selection of groups in this redshift range that biases the luminosity function towards bright galaxies. Excluding galaxies in groups with more than 8 members leads to more comparable results, though completeness remains slightly higher.

\section{Structures identification and density estimators}
\label{sec:structures}

The quantification of environment and the identification of structures can be done in many ways. These methods include Voronoi tesselations and Delaunay triangulation \citep[e.g.,][]{Marinoni+02}, weighted adaptive kernel smoothing, density at the radius of the n$^{th}$ neighbor \citep[e.g.,][]{Wang+20}, friends-of-friends (FoF) algorithms, and are based either on spectroscopic or photometric redshifts or the combination of both. Each method has advantages, such as the possibility to combine both photometric and spectroscopic measurements \citep[e.g., the Voronoi tesselations Monte-Carlo mapping -VMC- developed in][]{Lemaux+17, Lemaux+22, Hung+20, Hung+21}, or an easy implementation \citep[e.g., the FoF algorithm used in][]{Knobel+12, Iovino+16}, but most of those remain sensitive to the underlying completeness of redshift measurements and need extensive sets of simulations to tune the algorithm parameters. They also depend on the size of the underlying studied field.
\citet{Darvish+15} describe and compare several methods within the COSMOS field using photometric redshifts and show an overall agreement between those methods.
The goal of the present paper is not to further develop these methods but to have various estimates of environment and density. The methods used to quantify environment are described hereafter.

\subsection{Identification of structures using a FoF algorithm}
\label{sec:fof}

In order to identify structures within our MUSE data, we run a FoF algorithm, as done for the MUSE Hubble Deep Field South \citep[HDFS,][]{Bacon+15}.
FoF algorithms basically rely on two parameters that define in which volume friends are searched, one of which is the maximum transverse separation, based on angular separation measurements, and the other one is the maximum longitudinal separation, orthogonally to the plane of the sky, based on redshift measurements.
For the HDFS, no constraint was set on the transverse separation owing to the relatively narrow MUSE FoV. Since some fields are made of mosaics in the MAGIC sample, we use both transverse and longitudinal separations in our algorithm, as it is usually done for large scale spectroscopic surveys \citep[e.g.,][]{Knobel+09, Knobel+12, Iovino+16}.
Various definitions can be used for those separations.
For MAGIC, we use the transverse proper distance $\Delta d$ drawn from for the angular separation between sources and the angular diameter distance at the redshift $z$ of each source.
On the other hand, redshift differences combine expansion and actual relative motions between galaxies (see discussion in Sect. \ref{sec:cgr84_2groups}).
In order to quantify how galaxies are dynamically bound to given underlying large scale structures and dark matter halos, we use the velocity difference $\Delta v$ to constrain the longitudinal separation.
This velocity difference is inferred from redshift differences $\Delta z$ between sources at the redshift $z$ of each source as $\Delta v = c \Delta z/(1+z)$.

In \citet{Knobel+12} and \citet{Iovino+16}, the parameters of the algorithm were tuned with simulations to mimic their target selection based on photometric measurements, and their spectroscopic redshift success rate. Our dataset does not suffer any pre-selection and has a good and homogeneous spectroscopic redshift completeness over the whole 0.2 to 1.5 redshift range (see Sect. \ref{sec:completeness}). In addition, the redshift uncertainty in our sample is low, $\lesssim 30$~\kms\ in terms of velocity for both emission and absorption line objects (see Sect. \ref{sec:zmeasurements}).
We therefore use a simpler scheme and run the FoF algorithm using all galaxies in the MAGIC catalog with a secure redshift (confidence flag larger than or equal to 2), without any magnitude restriction.
This choice allows us to infer group membership for both high-mass and low-mass galaxies.
Including the latter in the group finding algorithm also helps in refining the group parameters (Sect. \ref{sec:group_properties}) thanks to better statistics.
For each source, transverse proper distances and velocity differences to all other galaxies are computed.
Following \citet{Iovino+16} prescriptions, we use an iterative approach with FoF parameters depending on the number of galaxies in the groups and use maximum physical and velocity separations $\Delta d = 375$~kpc and $\Delta v = 500$~\kms\ for groups of seven or more members, $\Delta d = 375$~kpc and $\Delta v = 470$~\kms\ for groups of six members, $\Delta d = 350$~kpc and $\Delta v = 440$~\kms\ for groups of five members, $\Delta d = 300$~kpc and $\Delta v = 350$~\kms\ for groups of four members, $\Delta d = 275$~kpc and $\Delta v = 325$~\kms\ for groups of three members, and $\Delta d = 225$~kpc and $\Delta v = 300$~\kms\ for pairs. The velocity separations are inferred from \citet{Iovino+16} FoF parameters at $z=0.1$, providing more stringent limits than considering higher redshifts.
Friends of a source are those within the maximum transverse and longitudinal separations. All friends of a friend are then considered as members of the same structure.

A refined analysis about groups statistics is provided in Sect. \ref{sec:magic_groups_properties}. Our goal here is to show how robust are the most massive groups with respect to FoF parameters and galaxies magnitude.
We find 26 groups with at least eight members, all having $0.3<z<1.4$. Using more relaxed constraints (by $30$\%) on both transverse and longitudinal separations only adds one or two galaxies for 20\% of those groups. Conversely, using more stringent constraints (by $30$\%) slightly reduces the number of galaxies in some of those groups, and splits three of them in smaller sub-groups.
We also run the FoF algorithm using a magnitude limit of 24.5 in the $z^{++}$ band on 3\arcsec\ diameter apertures, where spectroscopic redshifts completeness in MAGIC is higher than 70\% per magnitude bin. This reduces the number of groups with at least eight members to 18. Nevertheless, 21 out the 26 aforementioned groups are identified with at least five members, and the five remaining groups have either three or four members (all at $z>0.8$).
We also do the same exercise using only galaxies brighter than $K_s=22.6$, to mimic the pre-selection done in \cite{Iovino+16} and only find 13 groups with eight or more members. However, similarly to the previous test, 18 out the 26 groups found with the whole galaxy catalog are identified with at least five members within this $K_s$ band limited catalog, six have either three or four members (all at $z>0.8$), and two are not found at all (previously identified as having eight and nine members).
We are therefore quite confident that our FoF algorithm finds actual groups.

\subsection{Deriving group properties}
\label{sec:group_properties}

Once groups are identified thanks to the previously described FoF algorithm, we derive their properties.
First, group richness is estimated as the number of galaxies it has within MAGIC fields. This is done despite the fact that structures may extend beyond those fields. The richness we compute is thus a lower limit. Groups centered at the edges of MUSE FoV may be more impacted, but this should not be the case for the massive targeted groups.
The group redshift $z_g$ is also computed as the mean redshift of those members.

The line-of-sight velocity dispersion of group members is an important property to determine a reliable group mass. Following \citet{Cucciati+10} prescription, we estimate this velocity dispersion using the gapper method \citep{Beers+90} since it is a robust estimator, in particular when the number of group members is as low as five, as is the case for some of the groups found in our dataset. The velocity dispersion is therefore computed as:
\begin{equation}
\sigma_g = \frac{\sqrt{\pi}}{N(N-1)} \sum^{N-1}_{i=1} i(N-i) (\Delta v_{i+1} - \Delta v_{i}) \text{~,}
 \label{eq:gapper}
\end{equation}
where the velocities along the line-of-sight of each member $\Delta v_i$ are sorted in ascending order. Those velocities are inferred in the frame of the group from the difference in redshift between each member ($z_i$) and the group ($z_g$) as:
\begin{equation}
 \Delta v_i = c\times \frac{z_i - z_g}{1 + z_g} \text{~.}
 \label{eq:indiv_v}
\end{equation}
As a sanity check, we also compute the velocity dispersion as the standard deviation of the velocity distribution in each group. As expected, we find that the gapper method gives larger estimates than standard deviation. The difference decreases from 20\% to 10\% when the richness increases from three to ten galaxies, with no noticeable difference for groups with more than 20 members and no trend with redshift.
The estimate of the velocity dispersion may be overestimated if several sub-components are considered. However, we do not expect this to happen given that the MUSE FoV typically covers half the size of massive groups (see Sect. \ref{sec:magic_groups_properties}).

The velocity dispersion is a good proxy for the mass content of groups and is also used to infer their radius. We follow the prescriptions of \citet{Lemaux+12}, based on earlier works \citep[][]{Carlberg+97, Biviano+06, Poggianti+09}, to compute $R_{200}$, the radius at which the group density is equal to 200 times the Universe critical density and $M_{\rm Vir}$, the Viral mass at the Virial radius $R_{\rm Vir} = R_{200} / 1.14$, as:
\begin{equation}
R_{200} = \frac{\sqrt{3} \sigma_g}{10 H(z_g)} \text{~,~and}
 \label{eq:R200}
\end{equation}
\begin{equation}
M_{\rm Vir} = \frac{3\sqrt{3} \sigma_g^3}{11.4 G H(z_g)} \text{~,}
\label{eq:mvir}
\end{equation}
where $H(z_g)$ is the Hubble parameter at the redshift $z_g$ of the group, and $G$ is the Newton's gravitational constant. These prescriptions assume that groups are virialized.

We also estimate the position of the group center as the unweighted barycenter of all members within the MUSE fields.
This position is biased by the position of MUSE fields and may be offset with respect to the real center position.
Other estimates of group center positions are provided for groups also identified (i) in the zCOSMOS 20k group catalog \citep{Knobel+12}, (ii) in the COSMOS Wall \citep{Iovino+16}\footnote{The COSMOS Wall catalog only covers the redshift range $0.69<z<0.79$ and a fraction of the COSMOS field.}, and (iii) in the latest COSMOS X-ray catalog \citep{Gozaliasl+19}.
The corresponding zCOSMOS 20k group is the one with the largest number of members within each MAGIC group, if any. For both COSMOS Wall and X-ray group catalogs, we match each MAGIC group to the closest group in those catalogs having a good agreement on redshift ($\Delta z /(1+z) < 0.005$). For the COSMOS Wall, we further impose a center separation smaller than 2\arcmin , since we expect that our fields are centered on such groups but want to ensure that no possible match is missed.
In practice, all six matching groups between MAGIC and COSMOS Wall catalogs are separated by less than 25\arcsec , corresponding to less than 170~kpc.
The separation between the X-ray peak and our center is lower than $\sim 1$\arcmin\ except for one group (MGr15, CGr23\footnote{MGr and CGr stand for MAGIC group and COSMOS group, respectively, see Table \ref{tab:group_catalog_5members}.}) for which the separation is close to 3\arcmin , corresponding to $\sim 880$~kpc. The Virial radius determined by \cite{Knobel+12} for this group is 876~kpc while their center is at $\sim 150$\arcsec ($\sim 750$~kpc) from the X-ray position, which makes this association plausible.
There remains however a possibility that some matched X-ray groups do not correspond to MAGIC groups, especially when separations are large.

Once the center is determined, we further compute for each group member the projected distance $\Delta r$ to the group center from the on-sky projected angular separation and the angular diameter distance at the redshift $z_g$ of the group.

\subsection{Local density estimator using the VMC method}
\label{sec:vmc}

In order to have a local density estimate which is not biased by the oversampling of spectroscopic redshifts inside the MAGIC fields observed with MUSE, we apply the VMC method detailed in \citet{Lemaux+18} and \cite{Hung+20}, using the probabilistic model defined in \citet{Lemaux+22}, to estimate the local density from both photometric and spectroscopic redshifts catalogs. The photometric redshift catalog is COSMOS2015 \citep{Laigle+16}, and we include both zCOSMOS \citep{Lilly+07} and VIMOS Ultra-Deep Survey \citep[VUDS,][]{LeFevre+15} full spectroscopic catalogs that provide a fairly homogeneous sampling over the COSMOS field.

All objects with IRAC1 magnitudes lower than 24.8 are used to generate the maps, in order to ensure that the sample used is roughly stellar mass limited.
The VMC method assigns a redshift to each galaxy within this magnitude limit in order to constitute one Monte-Carlo iteration.
Photometric redshifts are used for all galaxies with either no spectroscopic redshift or with an insecure spectroscopic redshift (confidence flag $< 2$).
For each source having a secure spectroscopic redshift (confidence flag $\ge 2$), for each iteration, a number is drawn from an uniform distribution ranging from 0 to 1. If this number is below a likelihood threshold that depends on the spectroscopic redshift confidence flag, the spectroscopic redshift value is retained \citep[details about the method and these thresholds are provided in Appendix A of][]{Lemaux+22}. Otherwise, the photometric redshift information is used.
For each source for which the photometric redshift is used, the redshift is randomly assigned by sampling its photometric redshift probability distribution function which is treated independently from that of neighboring galaxies.

For a given iteration, $\pm 3.75$~Mpc deep slices with respect to the fiducial redshift center with a step of 0.75~Mpc (proper distances) are produced from $0.2\le z \le 1.5$.
For each slice, a Delaunay triangulation is performed to generate Voronoi tessellations using the positions of objects that are within the velocity/redshift range of the slice. The density is inversely proportional to the area of each Voronoi tile. A grid of 75~kpc wide pixels is produced with the value of the density for each slice. One hundred Monte-Carlo iterations are performed and, for each pixel of the grid, the final density is computed as the median over the iterations.

\begin{figure}
 \includegraphics[width=\columnwidth]{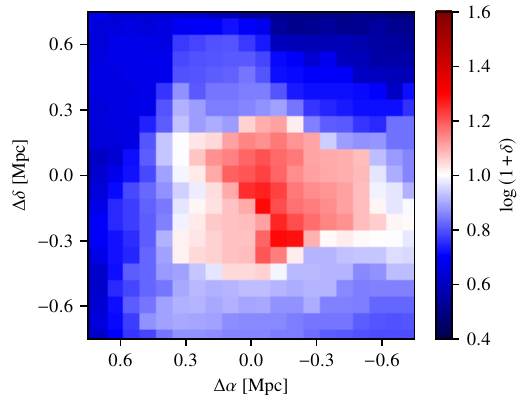}
 \caption{\label{fig:overdensity_cosmos}Map of the overdensity for CGr32 obtained with the VMC method. The displayed slice is at the mean redshift of CGr32 and covers the three fields mosaic.}
\end{figure}

In order to avoid edge effects where the data begin to fall off, a mask is applied to each slice to mask out all pixels where the density is less than 0.1 galaxy per Mpc$^2$.
The median density per slice is then calculated from the unmasked region of each slice.
Since the sampling might differ with redshift due (i) to the source density falling off naturally in an apparent magnitude limited sample to higher redshifts, and (ii) to variations of spectroscopic redshifts completeness, we compute the overdensity with respect to the median density $\Sigma_{\rm median}$ over the COSMOS field computed as $1+\delta = \Sigma / \Sigma_{\rm median}$ at each redshift slice, rather than the density $\Sigma$. Using the overdensity mitigates these effects to a large degree. Still, both density and overdensity are provided in our final catalog.
Each galaxy in our sample is thus attributed the density and overdensity of the closest pixel in the corresponding redshift slice (middle of the slice redshift range).
An example of the overdensity map at the redshift of CGr32 is provided in Fig. \ref{fig:overdensity_cosmos}.

\subsection{Local density estimator in structures using Voronoi tessellations}
\label{sec:muse_voronoi}

\begin{figure}
  \includegraphics[width=\columnwidth]{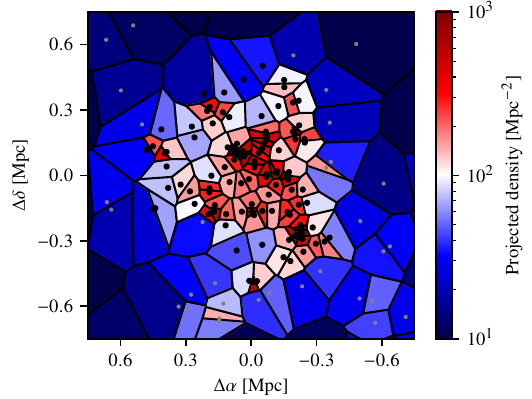}
 \caption{\label{fig:overdensity_voronoi}Map of the local density estimated from Voronoi tessellations for CGr32. Each galaxy is allotted a density equal to the inverse of the area of the associated Voronoi cell. Black dots correspond to galaxies of CGr32 that have a secure spectroscopic redshift in the MAGIC catalog, whereas gray dots correspond to galaxies from the \zCOSMOS\ catalog in the redshift range of CGr32.}
\end{figure}

In order to take advantage of our MUSE dataset and of its high spectroscopic redshifts completeness (see Sect. \ref{sec:completeness}), we compute an additional local density for each galaxy in any identified structure based on a similar but simpler approach than the above VMC method (Sect. \ref{sec:vmc}).
For each group, we perform a triangulation from the position of all galaxies with a secure spectroscopic redshift. Then Voronoi cells are allotted to each galaxy. Since each cell contains only one galaxy, the corresponding galaxy is attributed a density per unit area equals to the inverse of the Voronoi cell proper area in Mpc$^2$ at the redshift $z_g$ of the group.
This method suffers however from edge effects since tiles for galaxies at the edge of the group or of the MUSE FoV have undefined boundaries. To partially overcome this, we use the \zCOSMOS\ catalog (Khostovan et al. in prep, see Sect. \ref{sec:comparison_spectro_catalogues}) outside the MUSE FoV to add constraints on those 'edge' galaxies. We add all galaxies with a secure (confidence flag at least 2) spectroscopic redshift in the \zCOSMOS\ catalog within the total redshift range of the group, which boundaries are defined as the minimum and maximum redshifts of members identified in MUSE data.
An example of such a density map is shown for CGr32 in Fig. \ref{fig:overdensity_voronoi}. The overall density pattern displays a good qualitative match with respect to that of the VMC method in Fig. \ref{fig:overdensity_cosmos}.

The advantage of this method is that the density is computed using all secure group members rather than using redshift slices. However, apart from the edge limitation, it also only applies to galaxies in groups, even though it is still possible to infer upper limits for other galaxies. Last, we emphasize that we do not use any weighting scheme, which makes this density estimate sensitive to low-mass galaxies in the same way as for massive ones.

\subsection{Global density estimator in structures}
\label{sec:eta}

For any galaxy in a structure, we determine a global density estimator with the dimensionless parameter \citep{Noble+13}:
\begin{equation}
 \eta = \frac{|\Delta v|}{\sigma_g} \times \frac{\Delta r}{R_{200}}\text{~,}
 \label{eq:eta}
\end{equation}
where $\Delta v$ is the velocity of the galaxy within the group (see Eq. \ref{eq:indiv_v}) and $\Delta r$ is the projected distance to the center of the group.
This phase-space parametrization quantifies how galaxies are tied to the group they belong to, based on their normalized position and kinematics.
Using this parameter, galaxies can be separated between galaxies that recently started their infall onto the groups ($0.4 < \eta < 2$), backsplash galaxies that already passed once the pericenter of their orbits ($0.1 < \eta < 0.4$), and galaxies that have been accreted earlier ($\eta < 0.1$).
The limitation of this parameter is that it can only be defined in groups with well-defined centers, dispersion and size. It is therefore more reliable for groups with more than ten members.

\subsection{Analysis of MAGIC groups}
\label{sec:magic_groups_properties}

The MAGIC survey contains 76 pairs of galaxies and 67 groups, spread in 30 groups of three or four members, 18 groups of five to nine members, 13 groups of ten to 19 members, two groups of 20 to 29 members, three groups of 30 to 50 members, and one group of more than 50 members that can be considered as a cluster (CGr32) owing to both its richness and mass.
Depending on how defining the limit between groups and clusters, other groups, mostly those that were initially targeted, may be considered as low mass clusters.
Overall, the MAGIC sample covers a wide range of structure mass, from small groups of $\sim 10^{12}~M_\sun$ to clusters of a few $10^{14}~M_\sun$. For the sake of simplicity, we refer hereafter to a group any structure with at least three members.
We match 23, 6, and 11 groups over the 67 groups with three or more members to groups previously identified in the zCOSMOS 20k \citep{Knobel+12}, in the COSMOS Wall \citep[][]{Iovino+16} and in the X-ray \citep{Gozaliasl+19} group catalogs, respectively (see Sect. \ref{sec:group_properties}).
The 11 groups with an X-ray counterpart are among the 14 targeted massive groups. The three targeted groups with no X-ray detection all have \textit{XMM-Newton} depths of about $10^5$~s, though they have similar redshifts and dynamical masses as other groups with an associated X-ray signal. These non-detections are likely due to an intrinsically weaker X-ray emission probably induced by less hot gas in those systems. None of the other groups found in MAGIC have an X-ray counterpart, except eventually MGr55 (see Sect. \ref{sec:cgr84_2groups}). Indeed, they are less massive and their X-ray emission, if any, would be embedded in the emission of the most massive group in those fields, and would thus be difficult to disentangle.

\begin{sidewaystable*}
\caption{\label{tab:group_catalog_5members}MAGIC catalog of groups with more than 5 members.}
\centering
\begin{tabular}{cccccccccccccc}
\hline\hline
ID$_{\rm MAGIC}$ & R. A. & Dec. & $z_g$  & Richness & $\sigma_g$ & $\log{(M_{\rm Vir}/M_\sun)}$ & $R_{200}$ & ID$_{\rm COSMOS}$ & ID$_{\rm CW}$ & ID$_{\rm X-ray}$ & Sep    & $\log{(M_{200}/M_\sun)}$ & $R_{200,\rm X-ray}$ \\
         & J2000 & J2000 &        &       & km~s$^{-1}$ &        & kpc    &          &       &       & \arcsec &  & kpc \\
(1)      & (2) & (3) & (4)    & (5)   & (6)   & (7)    & (8)    & (9)      & (10)  & (11)  & (12)   & (13) & (14) \\
\hline
MGr3     & 150\degr 08\arcmin 24.4\arcsec  & 2\degr 04\arcmin 17.8\arcsec  &  0.221 &     5 &   312 &  13.62 &    692 & CGr44    &       &       &        &  &  \\
MGr12    & 149\degr 59\arcmin 00.6\arcsec  & 1\degr 47\arcmin 52.8\arcsec  &  0.339 &     9 &   389 &  13.87 &    808 & CGr51    &       &       &        &  &  \\
MGr15    & 149\degr 47\arcmin 20.4\arcsec  & 2\degr 10\arcmin 16.3\arcsec  &  0.355 &    10 &   129 &  12.43 &    266 & CGr23    &       & 20099 &    179 &  13.23 &    475 \\
MGr17    & 149\degr 43\arcmin 37.2\arcsec  & 1\degr 55\arcmin 07.7\arcsec  &  0.364 &    14 &   550 &  14.32 &   1126 & CGr61    &       &       &        &  &  \\
MGr23    & 150\degr 29\arcmin 33.0\arcsec  & 2\degr 04\arcmin 09.8\arcsec  &  0.440 &    15 &   547 &  14.29 &   1071 & CGr26    &       & 20088 &     19 &  13.61 &    612 \\
MGr28    & 149\degr 55\arcmin 44.8\arcsec  & 2\degr 31\arcmin 37.9\arcsec  &  0.470 &     6 &   157 &  12.66 &    302 &          &       &       &        &  &  \\
MGr31    & 149\degr 58\arcmin 52.7\arcsec  & 1\degr 48\arcmin 10.1\arcsec  &  0.480 &     5 &    75 &  11.69 &    143 & CGr975   &       &       &        &  &  \\
MGr34    & 149\degr 59\arcmin 40.9\arcsec  & 2\degr 15\arcmin 15.1\arcsec  &  0.496 &     7 &   192 &  12.91 &    364 & CGr149   &       &       &        &  &  \\
MGr35    & 150\degr 13\arcmin 32.5\arcsec  & 1\degr 48\arcmin 40.3\arcsec  &  0.528 &    11 &   393 &  13.84 &    731 & CGr28    &       & 20035 &     64 &  13.67 &    622 \\
MGr36    & 149\degr 49\arcmin 15.2\arcsec  & 1\degr 49\arcmin 17.8\arcsec  &  0.531 &    19 &   463 &  14.05 &    860 & CGr79    &       & 20039 &     14 &  13.60 &    589 \\
MGr45    & 149\degr 59\arcmin 50.3\arcsec  & 2\degr 15\arcmin 32.4\arcsec  &  0.659 &    12 &   319 &  13.53 &    549 & CGr114   &       & 20303 &     41 &  13.51 &    522 \\
MGr48    & 149\degr 55\arcmin 26.4\arcsec  & 2\degr 31\arcmin 59.5\arcsec  &  0.668 &     9 &   272 &  13.33 &    466 &          &       &       &        &  &  \\
MGr55    & 150\degr 03\arcmin 26.3\arcsec  & 2\degr 36\arcmin 32.8\arcsec  &  0.680 &    39 &   486 &  14.08 &    826 &          &       & 30231 &     66 &  13.70 &    605 \\
MGr57    & 150\degr 10\arcmin 16.7\arcsec  & 2\degr 31\arcmin 16.0\arcsec  &  0.696 &    22 &   374 &  13.73 &    629 & CGr172   & W5    & 10215 &      9 &  13.44 &    488 \\
MGr58    & 150\degr 03\arcmin 23.8\arcsec  & 2\degr 36\arcmin 08.6\arcsec  &  0.697 &    33 &   437 &  13.93 &    735 & CGr84    & W11   & 30231 &     42 &  13.70 &    611 \\
MGr60    & 149\degr 55\arcmin 49.1\arcsec  & 2\degr 31\arcmin 23.5\arcsec  &  0.698 &    12 &   201 &  12.92 &    338 & CGr173   &       &       &        &  &  \\
MGr63    & 150\degr 08\arcmin 30.5\arcsec  & 2\degr 04\arcmin 01.2\arcsec  &  0.725 &    44 &   404 &  13.83 &    669 & CGr30    & W4    & 20080 &     28 &  13.70 &    589 \\
MGr64    & 150\degr 01\arcmin 31.4\arcsec  & 2\degr 21\arcmin 21.6\arcsec  &  0.726 &    19 &   410 &  13.84 &    677 & CGr87    & W12   & 30324 &     17 &  13.37 &    457 \\
MGr65    & 149\degr 55\arcmin 19.2\arcsec  & 2\degr 31\arcmin 17.0\arcsec  &  0.730 &   107 &   928 &  14.91 &   1529 & CGr32    & W1    & 10220 &      4 &  14.37 &    983 \\
MGr66    & 150\degr 00\arcmin 20.5\arcsec  & 2\degr 27\arcmin 17.6\arcsec  &  0.731 &    24 &   427 &  13.89 &    703 & CGr35    & W3    & 20217 &      7 &  13.52 &    513 \\
MGr67    & 149\degr 51\arcmin 31.0\arcsec  & 2\degr 29\arcmin 29.0\arcsec  &  0.733 &    19 &   451 &  13.97 &    742 & CGr34    &       &       &        &  &  \\
MGr71    & 150\degr 02\arcmin 56.0\arcsec  & 2\degr 35\arcmin 32.3\arcsec  &  0.801 &     9 &   129 &  12.31 &    203 &          &       &       &        &  &  \\
MGr76    & 149\degr 55\arcmin 07.7\arcsec  & 2\degr 30\arcmin 44.3\arcsec  &  0.838 &    12 &   230 &  13.06 &    355 &          &       &       &        &  &  \\
MGr81    & 149\degr 55\arcmin 31.4\arcsec  & 2\degr 31\arcmin 35.0\arcsec  &  0.892 &     8 &    88 &  11.80 &    132 &          &       &       &        &  &  \\
MGr84    & 150\degr 10\arcmin 18.5\arcsec  & 2\degr 31\arcmin 25.3\arcsec  &  0.898 &    12 &   289 &  13.34 &    431 &          &       &       &        &  &  \\
MGr86    & 150\degr 01\arcmin 38.6\arcsec  & 2\degr 21\arcmin 11.2\arcsec  &  0.930 &     6 &   111 &  12.08 &    162 & CGr1444  &       &       &        &  &  \\
MGr88    & 150\degr 08\arcmin 46.3\arcsec  & 2\degr 04\arcmin 00.5\arcsec  &  0.938 &     8 &   166 &  12.61 &    242 &          &       &       &        &  &  \\
MGr90    & 150\degr 03\arcmin 29.2\arcsec  & 2\degr 36\arcmin 33.5\arcsec  &  0.943 &    12 &   223 &  12.99 &    324 & CGr285   &       &       &        &  &  \\
MGr101   & 150\degr 00\arcmin 29.5\arcsec  & 2\degr 26\arcmin 58.9\arcsec  &  1.097 &     5 &   228 &  12.99 &    303 &          &       &       &        &  &  \\
MGr107   & 149\degr 51\arcmin 33.1\arcsec  & 2\degr 29\arcmin 27.2\arcsec  &  1.171 &     8 &   139 &  12.32 &    177 &          &       &       &        &  &  \\
MGr108   & 149\degr 47\arcmin 28.7\arcsec  & 2\degr 10\arcmin 08.4\arcsec  &  1.171 &     5 &   131 &  12.25 &    168 &          &       &       &        &  &  \\
MGr112   & 149\degr 55\arcmin 17.4\arcsec  & 2\degr 31\arcmin 09.8\arcsec  &  1.193 &     5 &   148 &  12.40 &    187 &          &       &       &        &  &  \\
MGr113   & 149\degr 47\arcmin 43.4\arcsec  & 2\degr 10\arcmin 18.8\arcsec  &  1.244 &     6 &   175 &  12.60 &    214 &          &       &       &        &  &  \\
MGr117   & 150\degr 08\arcmin 36.2\arcsec  & 2\degr 03\arcmin 43.6\arcsec  &  1.275 &     6 &   292 &  13.26 &    351 &          &       &       &        &  &  \\
MGr120   & 149\degr 49\arcmin 23.9\arcsec  & 1\degr 49\arcmin 25.3\arcsec  &  1.308 &     9 &   222 &  12.90 &    262 &          &       &       &        &  &  \\
MGr122   & 150\degr 00\arcmin 29.5\arcsec  & 2\degr 27\arcmin 18.4\arcsec  &  1.320 &     5 &   180 &  12.62 &    211 &          &       &       &        &  &  \\
MGr124   & 150\degr 13\arcmin 38.3\arcsec  & 1\degr 48\arcmin 50.0\arcsec  &  1.372 &    11 &   149 &  12.37 &    170 &          &       &       &        &  &  \\
\hline
\end{tabular}
\tablefoot{(1) MAGIC group ID; (2) and (3) J2000 coordinates of group center; (4) Group redshift; (5) Number of group members; (6) Velocity dispersion (Gapper method); (7) and (8) Virial mass (log) and radius inferred from velocity dispersion; (9) zCOSMOS 20k group ID from \cite{Knobel+12}; (10) COSMOS Wall group ID from \cite{Iovino+16}; (11) X-ray conterpart ID from \cite{Gozaliasl+19}; (12) Separation between MAGIC and X-ray centers; (13) Mass inferred from X-ray (log); (14) Radius inferred from X-ray.}
\end{sidewaystable*}

The global group properties described in the previous sections are released in the MAGIC group catalog described in Appendix \ref{app:catalogs} (Table \ref{tab:group_catalog}), for all identified groups, including pairs. A version of this catalog for groups of more than five members is provided in Table \ref{tab:group_catalog_5members}. The local and global density estimates computed for each galaxy, as well as group membership, are provided in Appendix \ref{app:catalogs} (Table \ref{tab:galaxy_catalog}).
The MUSE FoV corresponds to about 400 to 450~kpc between $0.6<z< 0.8$, where most of the massive groups in MAGIC lie. Thus, the massive groups in MAGIC are typically twice as large as the MUSE field coverage, which means that we mostly survey the inner parts of groups within $R_{200}/2$. This may slightly underestimate their measured velocity dispersion, and may also increase the global fraction of quiescent galaxies in groups. We indeed expect galaxies infalling onto groups from their outer parts statistically have trajectories oriented close to the sky plane, hence with lower projected velocities within the group, and to be mostly star-forming.

\begin{figure}
 \includegraphics[width=\columnwidth]{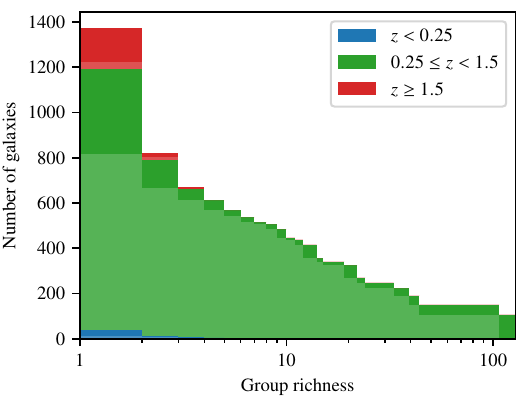}
  \caption{\label{fig:histo_richness}Histograms of the number of galaxies as a function of group richness in logarithmic scale for three redshift ranges.
  The height of the dark bins indicates the number of galaxies in groups of a given richness, whereas the  height of the bins of dark and lighter colors  corresponds to the number of galaxies in groups richer than the bin richness. The first bin corresponds to field galaxies and therefore contains all galaxies.}
\end{figure}

Figure \ref{fig:histo_richness} displays the number of galaxies in groups richer than a given number of members. Most groups and galaxies therein are in the redshift range of \oii\ emitters\footnote{All groups with a richness above five members are in this redshift range.}, whereas a non negligible fraction of field galaxies are at redshifts above $z=1.5$.
If we restrict to the 1154 galaxies with redshifts between 0.25 and 1.5, where the spectroscopic redshift completeness is the highest, we find that 778 galaxies belong to groups of two (galaxy pairs) or more members. We further find that 32.6\% are in the field, 28.7\% are in groups of nine or less members, 19.4\% are in groups of ten to 29 members, and 19.3\% are in groups of more than 30 members.

\begin{figure}
 \includegraphics[width=\columnwidth]{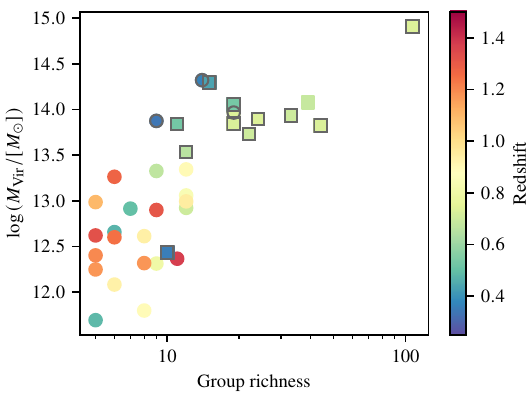}
 \caption{\label{fig:rich_mvir}Group richness as a function of their Virial mass for groups with five or more members with $0.25\le z<1.5$. Initially targeted groups have gray contours, and groups with X-ray counterparts are identified as colored squares. The color of the symbols refers to the redshift of each group.}
\end{figure}

\begin{figure}
 \includegraphics[width=\columnwidth]{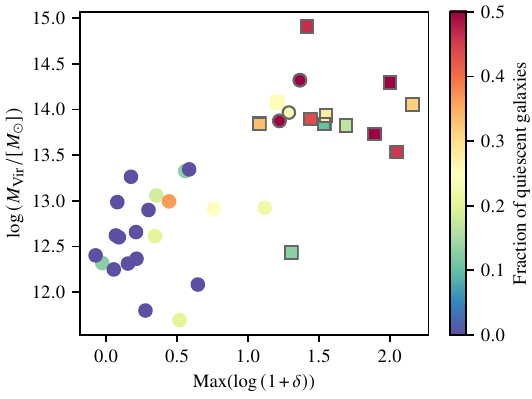}\\
 \caption{\label{fig:dens_mvir}Maximum of the VMC overdensity as a function of Virial mass for MAGIC groups with five or more members with $0.25\le z<1.5$. Initially targeted groups have gray contours, and groups with X-ray counterparts are identified as colored squares. The color of the symbols refers to the fraction of quiescent galaxies within each group.}
\end{figure}

There is a clear correlation between the richness of groups and their Virial mass estimated from dispersion (Eq. \ref{eq:mvir}), as illustrated in Fig. \ref{fig:rich_mvir}. This figure also shows that groups above redshift $z\sim 1$ are mostly low-mass and low-richness. It may be that their richness is underestimated due to their lowest completeness at low mass compared to groups at $z<1$.
We further investigate the link between the group mass, the fraction of quiescent galaxies and the density.
The population of quiescent galaxies is inferred from a color-color diagram as detailed hereafter in Sect. \ref{sec:red_sequence}.
The density can be estimated either from the VMC method, e.g., using the largest overdensity in the group, from the Voronoi tessellations method, e.g., using the 90$^{\rm th}$ percentile of the Voronoi density in the group, or $\Sigma_5$, the projected galaxy number density within a circle of radius equal to the distance to the fifth closest member to the group center \citep[similar to the definition of][]{Wang+20}. Those three density estimators provide fairly comparable results, and we show the results obtained with the VMC method in Fig. \ref{fig:dens_mvir}. This figure shows that massive groups are also those with the densest regions, though the most massive structure is not the one with the highest density.
Over the 15 most massive groups (Virial mass $\log{(M_{\rm Vir}/[M_\sun])} > 13.5$), 13 are the main groups we targeted, and one among the two groups that were not targeted has only five members, making the estimate of its dynamical mass less robust. Only one targeted group (CGr23, $z=0.355$) has a lower dynamical mass. This may be due to the fact that this field was slightly off-centered with respect to the actual center of the group owing to our initial observing strategy that aimed at optimizing the density of star-forming galaxies within one MUSE field (see Sect. \ref{sec:target_selection}). Indeed, the center determined by \cite{Knobel+12} is $\sim 30$\arcsec\ ($\sim 150$~kpc) away from the center derived from the MAGIC dataset.
\cite{Knobel+12} further inferred a velocity dispersion of 367~\kms\ and a dynamical mass of $\log{(M_{\rm Vir}/[M_\sun])} > 14$ for this group.
The velocity dispersion we measure ($\sigma_g=129$~\kms ) may thus be that of a sub-group of star-forming galaxies, and may underestimate both the radius and dynamical mass of the whole group, which could be the reason for the position of this group in Fig. \ref{fig:dens_mvir}.
The fraction of quiescent galaxies is also clearly larger for groups more massive than $\log{(M_{\rm Vir}/[M_\sun])} > 13.5$, where the local density is the highest (see further discussion in Sect. \ref{sec:red_sequence}).
The distribution of groups in these diagrams is due to combination of targeting massive groups and of searching blindly other structures within the MUSE FoV.
Indeed, blind searches lead to a higher probability to find structures with less than ten members than massive groups.
We also find a trend (not shown here) that the mass of a group is correlated to the mass of its most massive members.

\section{Galaxy properties in the MAGIC sample}
\label{sec:properties}

The MAGIC sample is an ideal laboratory to study the impact of environment on galaxy properties thanks to its high spectroscopic completeness, its ability to estimate accurately the environment of galaxies down to low stellar masses, and the diversity of environments probed.
Our goal here is to explore how the general properties of galaxies change for populations in various environments.
In this section, we focus the analysis in the redshift range $0.25\le z<1.5$ and restrict the analysis to galaxies with $z^{++}_{\rm app} < 25.5$ to ensure a completeness of at least 50\% locally (see Sect. \ref{sec:completeness}).

\subsection{Red sequence and environment}
\label{sec:red_sequence}

\begin{figure}
 \includegraphics[width=\columnwidth]{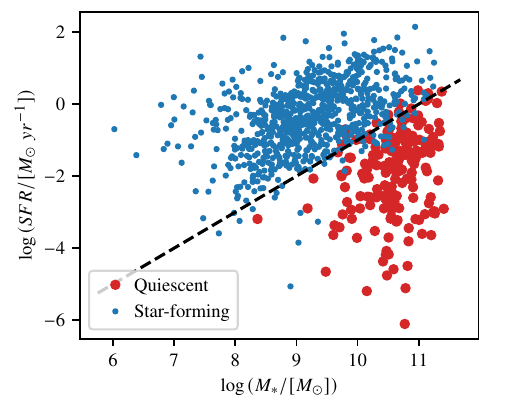}
  \caption{\label{fig:RS-MS}Location of galaxies with redshift $0.25\le z<1.5$ in the SFR vs $M_\star$ diagram (within 3\arcsec apertures). Blue and red dots correspond to star-forming and quiescent galaxies, respectively. The black dashed line corresponds to galaxies with $\log{(sSFR)}=-11$.}
\end{figure}

We identify 177 galaxies in the red sequence based on the $(NUV-r^+)-(r^+-J)$ two-color criterion defined in \citet{Ilbert+13} to select the quiescent population: $NUV-r^+>3.1$ and $NUV-r^+>3(r^+-J) +1$. Rest-frame colors are estimated from absolute magnitudes inferred from the \textsc{Cigale} run (see Sect. \ref{sec:sed_fitting}). This method has the advantage that it minimizes the effect of extinction and that it is less prone to evolution with redshift compared to a selection based on the specific star formation rate (sSFR, estimated in yr$^{-1}$). As a sanity check, we display in Fig. \ref{fig:RS-MS} the position of quiescent galaxies in $\log{(M_\star)}-\log{(SFR)}$ diagram showing that red-sequence galaxies have, as expected, lower SFR compared to galaxies on the main sequence, and that most quiescent galaxies have their sSFR lower than $10^{-11}$~yr$^{-1}$. The fraction of quiescent galaxies is 81\% among galaxies with $\log{(sSFR)}<-11$ and only 1\% with $\log{(sSFR)}\ge-11$.
Most of these galaxies have stellar masses above $10^9~M_\sun$, where completeness is above 70\% in MAGIC (Sect. \ref{sec:completeness}).

\begin{table}
\caption{\label{tab:fraction_redseq}Fraction of red sequence galaxies. Numbers in parentheses correspond to the total number of galaxies in each class.}
 \begin{tabular}{lcccc}
\hline\hline
  Classes & 1 &  2 &  3 &  4 \\
\hline
  VMC             & 5\% (499) & 23\% (200) & 29\% (154) & 47\% (129) \\
  Voronoi         & 5\% (474) & 15\% (117) & 32\% (195) & 37\% (196) \\
  Richness        & 6\% (308) &  9\% (274) & 33\% (211) & 34\% (189) \\
  $\sigma_g$      & 6\% (308) & 12\% (101) & 32\% (225) & 48\% (122) \\
  $\eta$          & 6\% (308) &  12\%  (84) & 32\% (176) & 41\% (188) \\
\hline
 \end{tabular}
\tablefoot{Classes correspond to increasing density from 1 to 4. For VMC, classes correspond to $1+\delta < 2$, $2 \le 1+\delta < 5$, $5 \le 1+\delta < 10$, and $1+\delta \ge 10$.
For Voronoi tessellations, the first class contains galaxies for which it was not possible to infer density (field and low richness).
For the other density parameters, the first class corresponds to galaxies in the field. Classes 2 to 4 correspond to
$\Sigma < 10~\textrm{Mpc}^{-2}$, $10~\textrm{Mpc}^{-2} \le \Sigma < 100~\textrm{Mpc}^{-2}$, and $\Sigma \ge 100~\textrm{Mpc}^{-2}$ for Voronoi tessellations,
$2\le N < 10$, $10\le N < 30$, and $N \ge 30$ for the richness,
$\sigma_g < 300$~\kms , $300~\textrm{\kms} \le \sigma_g < 500~\textrm{\kms}$, and $\sigma_g \ge 500~\textrm{\kms}$ for the velocity dispersion,
and $\eta \ge 0.4$, $0.1 \le \eta < 0.4$, and $\eta < 0.1$ for the phase-space parameter.
For $\sigma_g$ and $\eta$ parameters, galaxies in groups of nine or less members are further discarded.}
\end{table}

\begin{figure*}
 \includegraphics[width=\textwidth]{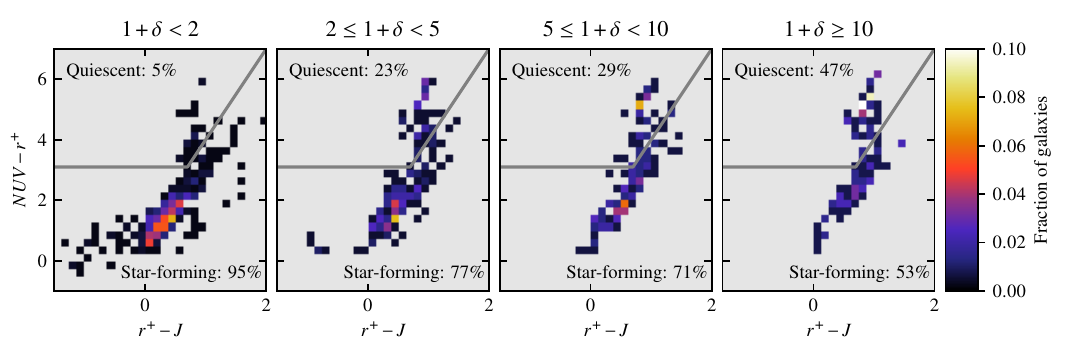}
 \caption{\label{fig:color-color_vmc}$NUV-r^+$ vs $r^+-J$ histograms for the VMC overdensity classes. The gray lines show the region of the diagram used to identify quiescent galaxies. The global fraction of quiescent and star-forming galaxies is provided in each quadrant.}
\end{figure*}

\begin{figure}
 \includegraphics[width=\columnwidth]{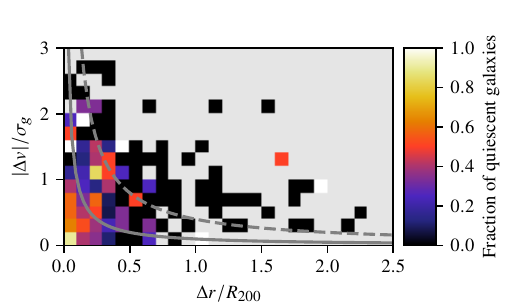}
 \caption{\label{fig:phasespace}Phase--space diagram showing the fraction of quiescent galaxies per bin for all groups with at least five members. The plain and dotted gray lines show the limits between the various classes at $\eta=0.1$ and $\eta=0.4$, respectively.}
\end{figure}

We further compute the fraction of galaxies in the red sequence as a function of environment in the redshift range of \oii\ emitters ($0.25\le z<1.5$). We split the sample in four classes as a function of galaxy environment inferred from various estimators, namely VMC overdentity, Voronoi tessellations, richness, group velocity dispersion and the phase-space parameter $\eta$.
Results are summarized in Table \ref{tab:fraction_redseq}, and we show in Fig. \ref{fig:color-color_vmc} the fraction of galaxies in the $(NUV-r^+)-(r^+-J)$ diagram for the four VMC overdensity $1+\delta$ classes. In this figure, we see that the fraction of quiescent galaxies gradually increases with overdensity and that galaxies also get redder on average for both colors. Similar trends are observed for other density estimators. Class 1 contains most of field galaxies for the VMC overdensity. It contains galaxies for which it was not possible to infer a density from the Voronoi tessellations method, including all galaxies in the field as well as some galaxies in groups with a low richness or at the edges of the fields.
For all other estimators, class 1 corresponds to field galaxies.
Considering the group richness, we see that the fraction of quiescent galaxies in groups of less than ten members remains relatively close to that of field galaxies. For this reason and because $\sigma_g$ and $\eta$ parameters are less reliable for groups with fewer galaxies, we further exclude galaxies in groups of nine or less members for those density estimators.
For the two classes of more than ten members, the fraction is similar. Among those two classes, we further see that groups with increasing velocity dispersion $\sigma_g$ host more quiescent galaxies. The position in the phase-space diagram as inferred from the $\eta$ parameter is independent of the group richness or velocity dispersion.
This estimator shows that the fraction of quiescent galaxies for recently accreted galaxies ($\eta \ge 0.4$, class 2) is low and only twice as large as for field galaxies and that it increases with the time spent in the structures, i.e. for backsplash galaxies (class 3) and galaxies that have been accreted earlier (class 4).
This is further illustrated in Fig. \ref{fig:phasespace} that displays the fraction of quiescent galaxies as a function of the position in the phase-space diagram ($|\Delta v|/\sigma_g$ vs $\Delta r/R_{200}$) for all groups of five and more members. Some bins at either high $|\Delta v|/\sigma_g$ or high $\Delta r/R_{200}$ show a fraction of 1, but this is due to low number statistics. Apart from those bins, the fraction of quiescent galaxies clearly peaks at both low $|\Delta v|/\sigma_g$ and $\Delta r/R_{200}$ values, and then decreases when those parameters increase. Our sample lacks galaxies at large group distances due to the limited MUSE FoV, except for the two mosaic fields (CGr32 and CGr84).

\begin{figure}
 \includegraphics[width=\columnwidth]{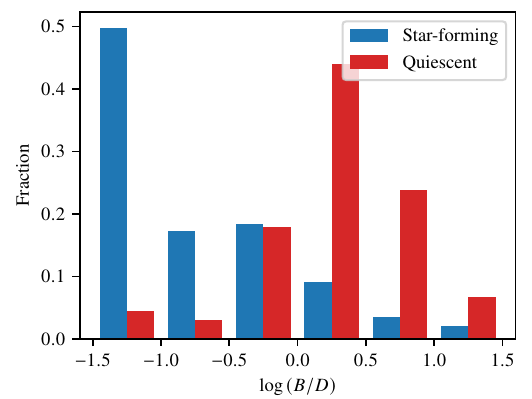}
 \caption{\label{fig:red_bulges}Normalized histograms of the bulge-to-disk ratio within the galaxy effective radius for star-forming and quiescent galaxies. The first and last bins contain all galaxies with B/D below 0.1 and above 10, respectively.}
\end{figure}

We also observe a clear dichotomy in morphology between quiescent and star-forming galaxies among the 732 galaxies that were modeled with a bulge-disk decomposition within the magnitude limit of $z^{++}\le 25.5$ in the \oii\ emitters redshift range. Star-forming galaxies are mostly disk-dominated (85\%  with $\log{B/D} < 0$), with 50\% being pure disks ($\log{B/D} < -1$), whereas 75\% of quiescent galaxies are bulge-dominated, as shown in Fig. \ref{fig:red_bulges}. This is in line with quiescent galaxies being mostly elliptical galaxies, as observed in the nearby Universe.

\subsection{Other galaxy properties}
\label{sec:galaxies_properties}

\begin{figure*}
 \includegraphics[width=\columnwidth]{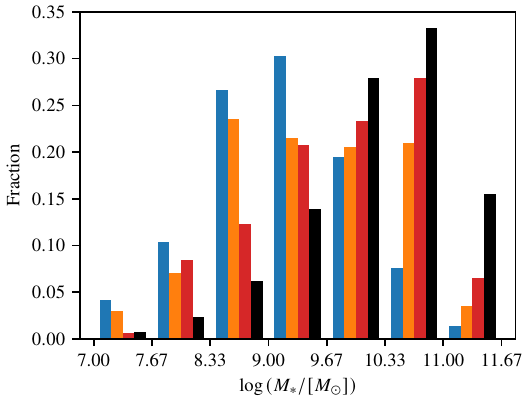}
 \includegraphics[width=\columnwidth]{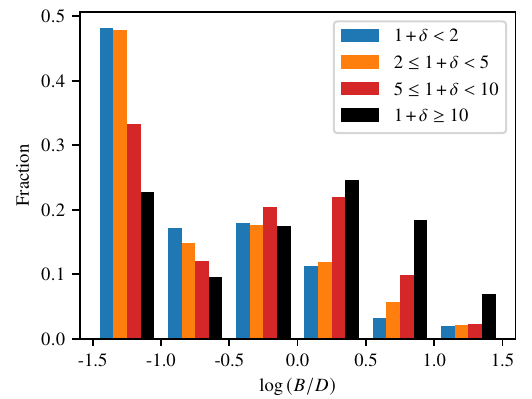} \\
 \includegraphics[width=\columnwidth]{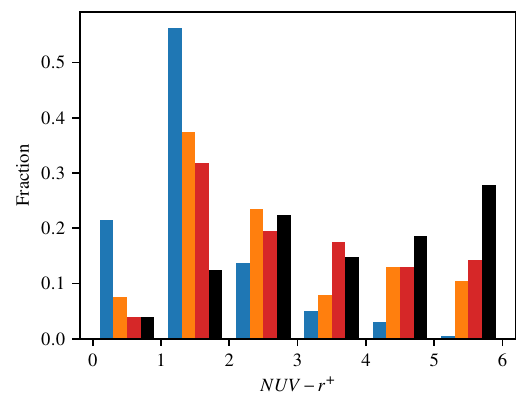}
 \includegraphics[width=\columnwidth]{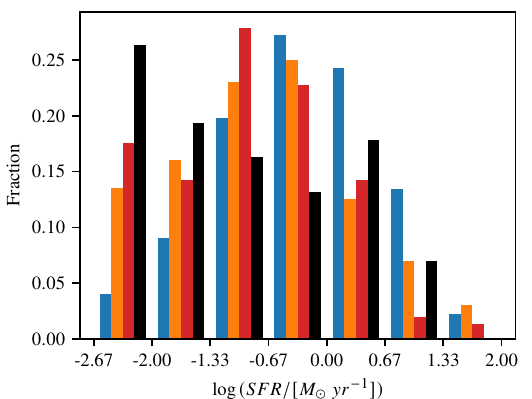}
 \caption{\label{fig:histograms1d}Histograms of stellar mass (top left panel), bulge-to-disc ratio within the galaxy effective radius (top right panel), color (bottom left panel), and SFR (bottom right panel) using the four VMC overdensity classes (see Sect. \ref{sec:red_sequence} and Table \ref{tab:fraction_redseq}). For each class in each histogram, the leftmost and rightmost bins contain all objects with values below and above the corresponding limits,  respectively.}
\end{figure*}

\begin{figure*}
 \includegraphics[width=\columnwidth]{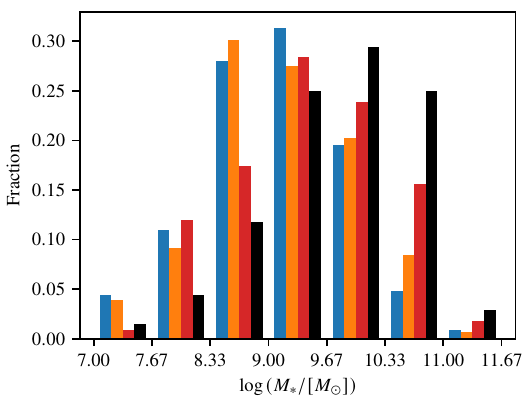}
 \includegraphics[width=\columnwidth]{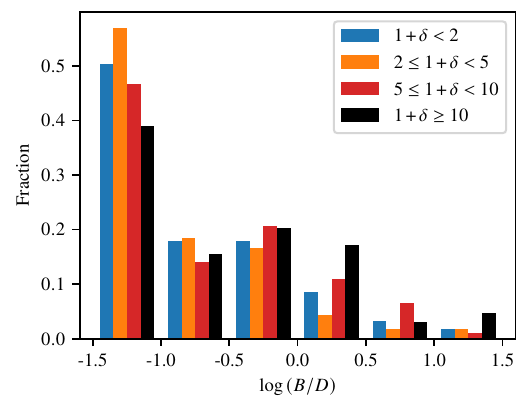}
 \caption{\label{fig:histograms1d_nored}Similar to Fig. \ref{fig:histograms1d}, but for the stellar mass and B/D only, and discarding quiescent galaxies.}
\end{figure*}

In order to estimate the impact of environment on galaxy properties such as stellar mass, color, SFR, and morphology, we split galaxies in the same four classes of increasing density as defined in Sect. \ref{sec:red_sequence} (see Table \ref{tab:fraction_redseq}).
We observe that galaxies get more massive, redder, less star-forming and more bulge dominated when the density increases, as illustrated in Fig. \ref{fig:histograms1d} by the histograms of those four quantities in four density bins using the VMC overdensity estimator.
These findings are partly due to the red sequence being more prominent in dense environments, as shown previously in Sect. \ref{sec:red_sequence}.
In Fig. \ref{fig:histograms1d_nored}, we show stellar mass (left panel) and B/D (right panel) histograms after further removing quiescent galaxies, which mostly reduces the fraction of galaxies in the highest stellar mass bins and with the highest B/D in the densest environments, compared to Fig. \ref{fig:histograms1d} where all galaxies are included.
Nevertheless, we still observe that the fraction of massive galaxies increases in dense environments (Fig. \ref{fig:histograms1d_nored}, left panel).
It results that star-forming galaxies appear both redder and slightly more star-forming in dense environments, owing to the color-mass and to SFR-mass relations for the main sequence of star-forming galaxies.
The latter is observed at large SFR when quiescent galaxies are also included, quiescent galaxies populating mostly the low SFR bins in Fig. \ref{fig:histograms1d}.
Last, the B/D distributions become quite similar whatever the environment when quiescent galaxies are discarded (Fig. \ref{fig:histograms1d_nored}, right panel), galaxies in dense environments having slightly more prominent bulges, probably related to the mass segregation.
This result, associated to the clear morphological dichotomy observed between disk-dominated star-forming galaxies and bulge-dominated quiescent galaxies and discussed in Sect. \ref{sec:red_sequence}, implies that the transition between the main sequence of star-forming galaxies and the red sequence must occur via a fast mechanism that makes galaxies redder by quenching star-formation and induces a change in the morphology at the same time.
Ram pressure stripping is a compelling mechanism since it is able to rapidly quench the activity of star 
formation of the perturbed galaxies on relatively short timescales \citep[$\lesssim$ 1 Gyr; e.g.,][]{Boselli+16, Boselli+22}.
However, being an hydrodynamic process, it can hardly affect the stellar component and produce the
prominent bulges observed in high density environments. It is thus likely that gravitational perturbations also contribute to the morphological transformation of these objects \citep[e.g.,][]{Boselli+06}.
Our results are in line we previous results obtained from spectroscopic surveys at similar redshifts \citep[e.g.,][who studied stellar mass functions as a function of a comparable local over-density estimator as the one used in our analysis]{Tomczak+17}.
Similar results are obtained with any other density parameters (richness, group velocity dispersion $\sigma_g$, Voronoi tessellations density, and phase-space parameter $\eta$).

It is clear that galaxy properties continuously change as environment gets denser, and the fact that the population of low-mass galaxies in groups is almost nonexistent may be due either (i) to the merger history, with low mass being merged before entering massive groups (e.g., during transit along the cosmic web, walls, and filaments) or (ii) to the dense environment itself, quenching star-formation of low-mass galaxies, and limiting our ability to assign a secure spectroscopic redshift for those galaxies without any emission line.
The impact of environment on the main sequence in the MAGIC sample has already been investigated in \citet{Mercier+22} who showed that at a given mass, the SFR is reduced by up to 0.15 dex in dense environments, which is too low to justify the drastic changes in galaxy properties we observe. There may however still exist a population of undetected intermediate to low-mass quiescent galaxies with no emission lines or low-mass galaxies on the main sequence with reduced star formation.

\section{Peculiar features found in MAGIC groups}
\label{sec:peculiar-features}

Apart from the galaxy and group catalogs, the MAGIC datacubes contain unique features.
Our goal here is not to be exhaustive but rather to highlight the most noticeable findings.

\subsection{An overdensity spanning CGr32, CGr34, CGr35 fields?}

CGr32, CGR34 and CGr35 fields are spatially close to each others and have their main identified group in the COSMOS Wall at $z=0.7304$, $z=0.7324$, and $z=0.7312$, respectively (see Fig. \ref{fig:cgr_location} and Tables \ref{tab:log} and \ref{tab:group_catalog_5members}).
On the one hand, the angular separation between CGr32 (MGr65) and CGr34 (MGr67) and between CGr32 and CGr35 (MGr66) is only of 4.2\arcmin\ ($\sim 1.8$~Mpc at $z=0.730$) and 6.4\arcmin\ ($\sim 2.8$~Mpc at $z=0.730$) respectively, which is less than about twice the radius $R_{200}\sim 1.5$~Mpc of CGr32. On the other hand, the scaled redshift separation between those groups is $\Delta z /(1+z)=0.00116$, corresponding to only $\sim 350$\kms , and $\Delta z /(1+z)=0.00046$, corresponding to only $\sim 140$\kms , respectively. The redshift distribution of galaxies in CGr32 includes those of both CGr34 and CGr35.
In the 20k zCOSMOS group catalog of \citet{Knobel+12}, some galaxies, mainly those with photometric redshifts only, have non null probabilities to be part of several of these groups, due to the method used to assign these galaxies to groups and to the large uncertainty on their actual redshift. Thanks to our MUSE data, we are able to assign a secure spectroscopic redshift to a significant fraction of these galaxies in the MAGIC fields.
We find no photometric candidate of CGr32 in the CGr34 or CGr35 fields, no candidate of CGr35 in the CGr32 or CGr34 fields, and no candidate of CGr34 in the CGr35 field. But several photometric candidates of CGr34 are in the CGr32 field. All of these have a higher probability to be part of CGr32. This raises the question about CGr32, CGr34 and CGr35 being one unique structure. CGr34 and CGr35 could also be either off-centered overdensities of the CGr32 cluster or groups infalling along filaments onto the CGr32 cluster, which seems to be supported by the filamentary structure of the COSMOS Wall \citep{Iovino+16, DarraghFord+19}. A wider field coverage with MUSE would be needed to clearly distinguish those options.

\subsection{Close groups in CGr84}
\label{sec:cgr84_2groups}

\begin{figure}
 \includegraphics[width=\columnwidth]{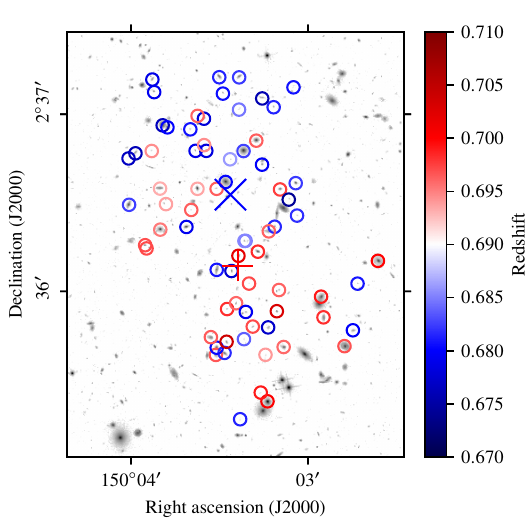}\\
 \includegraphics[width=\columnwidth]{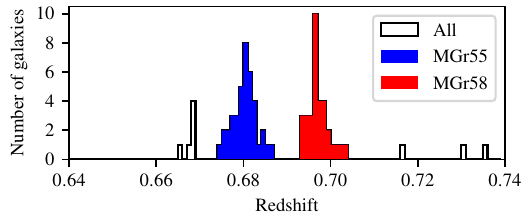}
 \caption{\label{fig:cgr84_2groups}Two main groups in the CGr84 mosaic identified with the FoF algorithm. The top map shows the F814W HST-ACS image in logarithmic scale (arbitrary unit) covering the two MUSE fields over CGr84. Colored circle correspond to galaxies with MUSE secure spectroscopic redshifts (confidence $\ge 2$) in the range of $0.67<z<0.71$. The redshift distribution is shown on the bottom histograms. Two structures in blue (MGr55, 39 members) and red (MGr58 or CGr84, 33 members) are identified and clearly separated in redshift space, though they overlap in their spatial distribution. Their respective centers from MUSE data are displayed as blue and red crosses.}
\end{figure}

The CGr84 mosaic exhibits two groups, MGr55 and MGr58 (CGr84) close in redshift space (see Fig. \ref{fig:cgr84_2groups} and Table \ref{tab:group_catalog_5members}) at $z=0.68031$ and $z=0.69711$ respectively and both quite rich (39 and 33 members respectively).
To assess whether those two groups may be gravitationally linked, we estimate on the one hand that their velocity difference is $\sim 2980$~\kms\ based on their scaled redshift difference $\Delta z /(1+z)=0.00995$, and on the other hand that their proper longitudinal separation is $\sim29$~Mpc if the redshift difference is only due to expansion.
The velocity dispersions of both structures are similar and slightly lower than 500~\kms, which corresponds to a Virial radius of about 1~Mpc and a halo mass of about $10^{14}~M_\sun$.
Assuming that the velocity difference is not due to expansion, having the two groups gravitationally bound would require their actual separation to be lower than about 120~kpc so that their velocity difference gets lower than the escape velocity. This is much smaller than both their Virial radii and their transverse extent and is thus unlikely.
On the other hand, structures with such masses and proper longitudinal separation cannot be bound because the Universe expands faster than the escape speed at a distance of 29~Mpc, which is lower than 200~\kms.
Another alternative would be that the two groups are located within a unique dark matter halo of much larger mass and scale, but such a scenario is unlikely. These groups are therefore most likely two distinct structures along the line-of-sight.

Interestingly, MGr55 was not found in the zCOSMOS 20k group catalog \citet{Knobel+12}, due to galaxies in this group being not observed in spectroscopy with VIMOS probably due to a combination of the a priori selection and of the moderate spatial sampling rate of the zCOSMOS 20k survey. This group also felt just outside the redshift range investigated by the COSMOS Wall survey \citep{Iovino+16}. An X-ray counterpart is also identified in \cite{Gozaliasl+19}, but due to the spatial and redshift proximity of the two groups, it could be associated to either one or the other group. For this reason, we have associated this X-ray group named W11 to both MUSE groups in Table \ref{tab:group_catalog_5members}.

We further searched for those structures in the VMC mapping. Overdensities are found at the redshift of both groups and their spatial distribution is consistent with the offset we observe. However, whereas these maps are in practice good indicators of (over)density as shown in previous sections, the overdensity does not drop sharply between the redshift of the two groups MGr55 and MGr58, but rather smoothly evolves,
whereas our MUSE data show that the structures are clearly separated in redshift space.
This highlights the advantages of using no a priori selection function, of making no assumption based on photometric redshifts, and of having a high spectrocopic redshifts completeness.

\subsection{Extended ionized gas nebulae in CGr172}
\label{sec:extended-nebulae}

As part of the MAGIC survey, two large ionized gas nebulae associated to massive groups have been investigated in detail. In the dense group CGr30 ($M_{\rm Vir}\sim 5-7\times 10^{13}~M_\sun$), a gigantic ($\sim 100$~kpc wide) nebula associated to interactions between galaxies, including one with an active galactic nucleus (AGN), and intragroup gas was found at $z\sim 0.7$. \cite{Epinat+18} suggested that a large disk may be reforming around a massive galaxy and the gas is ionized by a combination of electro-magnetic radiations from young stars, AGN and shocks. On the other hand, \cite{Boselli+19} showed the first evidence for ram-pressure stripping at $z\sim 0.7$ in two galaxies with $\sim 100$~kpc long tails of \oii -bright ionized gas, crossing CGr32, the most massive group of MAGIC ($M_{\rm Vir}\sim 2-8\times 10^{14}~M_\sun$).
These nebulae resemble the jellyfish galaxies identified in the GASP survey within local massive clusters \citep[e.g.,][]{Poggianti+17} or perturbed systems in clusters at intermediate redshifts ($0.3<z< 0.5$) observed with MUSE by \cite{Moretti+22}, where the availability of a comprehensive set of emission lines allowed the identification of photo-ionization as the dominant ionizing mechanism.

\begin{figure}
 \includegraphics[width=\columnwidth]{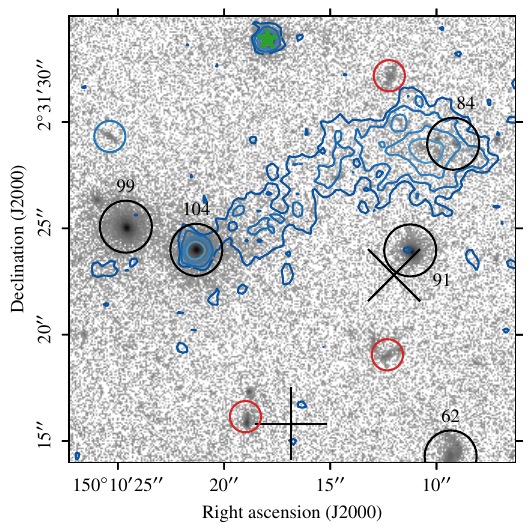}
 \caption{\label{fig:cgr172_map}F814W HST-ACS image of the filamentary structure found in CGr172 at $z\sim 0.696$ in logarithmic scale (arbitrary unit). The circles correspond to galaxies in the COSMOS2015 catalog within the field with secure redshifts, either within the same group as the extended emission (black), in the foreground (blue), or in the background (red). The green star marks a foreground star and the emission around it is an artifact. The black plus symbol and cross show the position of the group barycenter and X-ray center, respectively. The IDs of galaxies within CGr172 are indicated. The blue contours represent the \oii\ doublet in emission, at levels 0.25, 0.5, 1, and $2\times 10^{-17}$~erg~s$^{-1}$~cm$^{-2}$~arcsec$^{-2}$, after a Gaussian spatial smoothing of two pixels FWHM. Contours are estimated by collapsing the MUSE cube over 20~\AA\ at the observed wavelength of both lines, and by subtracting the continuum using 50~\AA\ wide adjacent wavelength ranges on both sides of the lines.}
\end{figure}

We report here the finding of a $\sim100$~kpc bridge of ionized gas seen in \oii\ between two members of the group CGr172 (MGr57) at $z\sim 0.69648$. This group contains 22 members and has a mass of $M_{\rm Vir}\sim 2-5\times 10^{13}~M_\sun$. We show in Fig. \ref{fig:cgr172_map} the F814W HST-ACS image as well as the contours of the \oii\ doublet emission at the redshift of the group.
This ionized gas nebula has a tail shape that connects galaxies 104\_CGr172 on the eastern side, to 84\_CGr172 on the western side, separated by a distance of 93~kpc and having a velocity difference of about 126~\kms .
The latter is a diffuse low-mass ($M_\star=8\times 10^{8}~M_\sun$) and star-forming ($SFR=1.5~M_\sun~\textrm{yr}^{-1}$) galaxy, composed of several clumps in the HST image. It is characterized by a high activity of star formation for its stellar mass which puts it above the main sequence relation. It seems that more small clumps are embedded in the tail between the two objects.
Interestingly, 104\_CGr172 appears to be in pair with 99\_CGr172, with a projected proper separation of 24~kpc and a velocity difference of about 154~\kms . These two galaxies are both on the red sequence and are the third and fourth most massive galaxies of the group, with stellar masses of $M_\star=1.3\times 10^{11} M_\sun$ (104\_CGr172) and $M_\star=1.1\times 10^{11} M_\sun$ (99\_CGr172). The group center is close to 104\_CGr172, both using the barycenter and the X-ray estimates.

At a first glance, the ionized gas tail might suggest gas removed during a ram-pressure stripping event from the massive galaxy 104\_CGr172, especially because the tail occurs close to the group center where hot intragroup gas is expected from the X-ray detection.
However, the presence of star-forming clumps far out in the tail, close and within 84\_CGr172, may indicate that the tail is the result of tidal interactions.
Using the formalism of \citet{Henriksen+96} as suggested in \citet{Boselli+22}, we can quantify the acceleration exerted by 104\_CGr172 on 84\_CGr172 and compare it to the one keeping matter linked to the gravitational potential well of 84\_CGr172.
Assuming a radius of 4.9~kpc (0.5\arcsec) for 84\_CGr172, corresponding to twice its effective radius, we estimate that the radial acceleration exerted by 104\_CGr172 on 84\_CGr172 is 20 times smaller than the acceleration keeping matter anchored to the gravitational potential well of 84\_CGr172. Assuming that the gas extent within 84\_CGr172 is four times its effective radius (1\arcsec, 9.8~kpc), the ratio between the accelerations reduces to only one third. We can thus not exclude that the observed extended tail associated to 104\_CGr172 is composed of ionized gas and stars removed from the very active object 84\_CGr172, especially if those two objects were closer at an earlier stage of the interaction.
During this kind of interaction, the different components of the baryonic matter can be removed from the perturbed system outside a truncation radius, while the one located in the inner regions can infall within the nucleus, sometimes boosting locally the activity of star formation. The clumps observed close to 84\_CGr172 can thus result from the turbulent stripped material, while the central ionized gas of 104\_CGr172 can be due to gas infalling onto the inner regions.
We made the same computations to quantify the tidal perturbation of 99\_CGr172 on 104\_CGr172 and found that such a tidal perturbation is unlikely, with a ratio of acceleration lower than 1/30, owing to the high mass and compact size (effective radius of about 2~kpc) of 104\_CGr172, unless those two galaxies were closer in the past. In that case, the groups of clumps could be viewed as a group of tidal dwarf galaxies.
Another scenario is that, rather than 104\_CGr172, 84\_CGr172 is suffering ram-pressure stripping by crossing the halo of 104\_CGr172 close to the group center, with stripped gas infalling on 104\_CGr172 and inducing a burst in its central regions. There could also be a combined effect of ram-pressure and tidal stripping exerted by the halo of 104\_CGr172 on 84\_CGr172, such as what happens to local group dwarfs in the halo of the Milky-Way \citep{Mayer+06}.
There is no evidence for AGN activity in the spectrum of 104\_CGr172, which rules out a priori the possibility of this feature to be an ejecta induced by nucleus activity.

Our goal here is only to highlight one of the most impressive nebulae found within the MAGIC dataset. A further analysis would consist in performing a systematic search of galaxies with either ram-pressure stripping or tidal stripping tails in MAGIC datacubes. This would allow one to draw statistics on the mechanisms at play to efficiently remove gas in galaxies, from an emission line flux-limited survey spanning a variety of groups.

\subsection{Quasar illuminated nebula in CGr84}
\label{sec:qso}

\begin{figure}
 \includegraphics[width=\columnwidth]{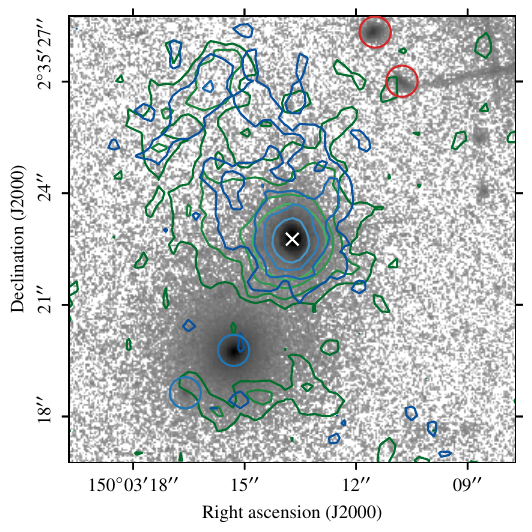}
 \caption{\label{fig:qso_map}F814W HST-ACS image of the surroundings of the QSO 231\_CGr84 at $z=0.7002$, centered on the QSO (identified with a white cross), in logarithmic scale (arbitrary unit). The circles correspond to galaxies in the COSMOS2015 catalog within the field with secure redshifts,
 either in the foreground (blue) or in the background (red).
 The green and blue contours represent the \oiiia\ and \oii\ emission lines respectively, at levels 0.25, 0.5, 1, and $2\times 10^{-17}$~erg~s$^{-1}$~cm$^{-2}$~arcsec$^{-2}$ after a Gaussian spatial smoothing of two pixels FWHM. Contours are estimated by collapsing the MUSE cube over 20~\AA\ at the observed wavelength of both lines, and by subtracting the continuum using 50~\AA\ wide adjacent wavelength ranges on both sides of the lines.}
\end{figure}

A quasi stellar object (QSO) is located within field CGr84-M1 at a redshift $z = 0.7002$, in group CGr84 (MGr58, $z=0.69711$), corresponding to the source 231\_CGr84. This QSO was previously reported in \cite{Prescott+06} as SDSS~J100012.91+023522, and in \cite{Heintz+16} as Q~100012.9+023522.8.
This galaxy has a stellar mass inferred from SED fitting of $2.6\times 10^{10}~M_\sun$, and it appears star-forming though strongly bulge-dominated.
However, these properties are not reliable because the QSO source biases both the morphology and the spectrum of the host galaxy.
It has a velocity of 545~\kms\ within the group, larger than the dispersion of CGr84 ($\sigma_g=437$~\kms ), and is located at a distance of 336~kpc from the center of the group, which is about half $R_{200}=735$~kpc of CGr84. The probability for this galaxy to be the brightest central galaxy of the group is thus low.
The QSO spectrum displays bright and broad emission lines and is much brighter in \oiii , \hb , or \mgii\ than in \oii .
It is surrounded by an ionized gas nebula shining in \oii , but brighter and more extended in \oiiia\ (see Fig. \ref{fig:qso_map}), which suggests that the surrounding gas is ionized by the QSO.
The nebula covers almost 10\arcsec\ from South to North ($\sim 70$~kpc in proper distance at the redshift of the QSO). It mostly extends to the North-East\footnote{A faint galaxy not identified in the COSMOS2015 photometric catalog is found from blind search at the redshift of the group at this location, probably due to the presence of the nebula. This may be a fake source.}, while there is a gap between the southernmost part of the nebula and the QSO. This gap may be due to 213\_CGr84, a bright and massive foreground galaxy at a redshift $z=0.66824$ bridging this gap, that is thus not a member of CGr84. Indeed, this galaxy may induce noise in the data and partially screen the nebula behind it.
The nebula is therefore not associated to any bright galaxy except the QSO itself.
The corresponding X-ray group W11 \citep{Gozaliasl+19} may actually be partly associated to this QSO.

Owing to its velocity within the group, this galaxy is likely traveling at high speed within the hot intragroup medium, detected in X-ray.
The elongated and asymmetric shape of the northern part of the nebula might therefore be formed by the external pressure acting on the ionized gas.
Indeed, the orientation of the nebula is in relative agreement with the direction of the group center (red cross in top panel of Fig. \ref{fig:cgr84_2groups}, where the QSO is the southernmost galaxy circled in red).
It has been claimed that a nuclear activity can be triggered during a ram pressure stripping process \citep{Poggianti+17, Peluso+22}, and such an association has also been observed around QSO at intermediate redshifts \citep{Johnson+18, Helton+21}.
This result, however, has been recently questioned by \cite{Boselli+22} and \cite{Cattorini+23} who did not find any increase in the nuclear activity in local galaxies suffering a ram pressure stripping event.
Large nebulae associated to ram pressure stripping are also observed in the most massive structure of MAGIC, CGr32, but not associated to nuclear activity \citep{Boselli+19}. On the other hand, in CGr30, an extended nebula is associated to an AGN, but in this case, interactions between galaxies seems more probable \citep{Epinat+18}, though ram pressure stripping could also be acting.
It is also possible that the gas seen in emission around the QSO in CGr84 originates from the circum-galactic or from the intragroup media.

We estimate the monochromatic luminosity at 5100~\AA\ by fitting the \hb\ line in the QSO spectrum, following the method described in \citet{Shen+11}. We further apply a bolometric correction \citep{Richards+06} to infer a bolometric luminosity of $\log{(L_{\rm bol}/[\textrm{erg}~\textrm{s}^{-1}])}=45.9$.
The \hb\ FWHM based on a joint narrow plus broad line fit including a power-law continuum and iron complex template from \citet{Boroson+92} is $FWHM_{H\beta} = 2200$ \kms .
Using the single-epoch virial estimator from \citet{Shen+11}, this results in an estimated super massive black hole mass (SMBH) mass of $\log{(M_{\rm BH}/[M_\sun])}=8.0$ which would mean that the QSO is accreting at a rate of $L=0.6 L_{\rm Eddington}$.

The giant, optically emitting nebula presented here is found around a QSO that is lower in both bolometric and NUV luminosities, and in SMBH mass by a factor of about ten relative to other QSO at intermediate redshifts with published nebulae on similar scales \citep{Johnson+18, Helton+21, Johnson+22, Chen+23, Liu+24}. This suggests that such giant nebulae might be quite common around more typical quasars.

\subsection{Gravitational arc in CGr32}
\label{sec:arc}

\begin{figure}
 \includegraphics[width=\columnwidth]{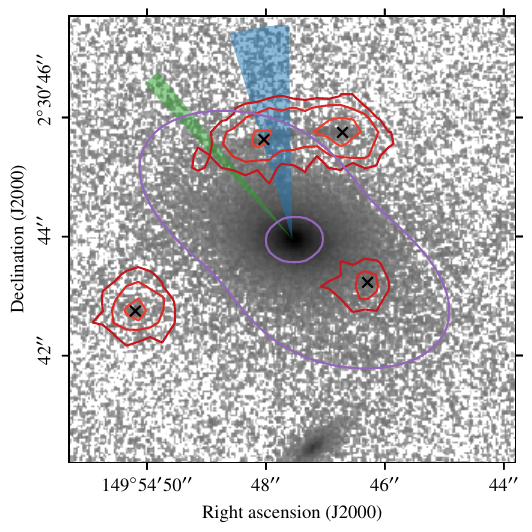}
 \caption{\label{fig:arc_map}F814W HST-ACS image centered on the lensing source, 152\_CGr32, at $z=0.72519$ in logarithmic scale (arbitrary unit). The red contours represent the \lya\ emission of the lensed source at $z=4.0973$ at levels 1, 2, and $4\times 10^{-17}$~erg~s$^{-1}$~cm$^{-2}$~arcsec$^{-2}$.
 It is estimated by collapsing the MUSE cube over 20~\AA\ at the observed wavelength of the \lya\ line, and by subtracting the continuum using 50~\AA\ wide adjacent wavelength ranges on both sides of the \lya\ range. The four images are identified with black crosses, the lens model critical lines are displayed in purple,
 and the blue wedge indicates the range in the directions to the massive structure responsible for the shear in the lens model, at 1$\sigma$. The green wedge represents the center directions between the barycenter of CGr32 members inferred from MUSE data and the X-ray center.}
\end{figure}

A gravitational arc (514\_CGr32) is clearly detected in the field CGr32-M3 around the galaxy 152\_CGr32 at $z=0.72519$. This galaxy is passive and is the most massive of cluster CGr32 (MGr65) with a stellar mass of $2.57\pm 0.13 \times 10^{11}~M_\sun$.
This arc was already identified by \citet{Faure+08} as COSMOS5939+3044 and is composed of two elongated images of the same galaxy North of 152\_CGr32. Thanks to the unambiguous detection of the asymmetric \lya\ line with MUSE, we are able to determine that the arc source is at $z=4.0973$, and to find two other images of this galaxy at the same redshift (513\_CGr32 and 530\_CGr32) on the other side of the lens, as illustrated in Fig. \ref{fig:arc_map}.
The large angular separation between the lensed images indicates the presence of massive structure in addition to the lens.
We model the lensing configuration of this system using a single isothermal elliptical potential at the location of 152\_CGr32 (parameterized by its velocity dispersion $\sigma$, ellipticity $e$ and angle $\theta$), together with an external shear (parameterized with an amplitude $\gamma_{\rm ext}$ and shear angle $\theta_{\rm ext}$). We are able to recover the locations of the four lensed images with an rms of 0.04\arcsec . The velocity dispersion of the main lens is $290\pm 15$~\kms , and we find a significant shear $\gamma_{\rm ext}=0.27\pm 0.08$ and $\theta_{\rm ext}=-81\pm 8$\degr .
The value of $\sigma$ found can be compared with the one inferred from the MUSE spectrum. We run the penalized pixel-fitting (pPXF) code\footnote{\url{http://www-astro.physics.ox.ac.uk/~mxc/software/}} \citep{Cappellari+04} on the lens integrated spectrum to fit its stellar continuum and infer its velocity dispersion.
The spectrum is extracted over a 6\arcsec\ wide square aperture. We use the MILES stellar population synthesis library \citep{Sanchez-Blazquez+06, Vazdekis+10, Falcon-Barroso+11} and take into account the lower spectral resolution ($FWHM=2.5$~\AA ) of this stellar library, compared to MUSE, taking advantage of the intrinsically large dispersion of the galaxy, to estimate that the velocity dispersion of the lens is about 250~\kms .
This value shows a good agreement with the results from our lens model. The orientation of the shear hints to the presence of a very massive structure at a position angle of $\sim 9\pm8$\degr\ (blue wedge in Fig. \ref{fig:arc_map}). This direction is in fair agreement of about $\sim 20-25$\degr\ at 1-$\sigma$ with the direction of CGr32 center found in X-ray and from MUSE spectroscopic redshifts (39\degr\ and 44\degr\ respectively, green wedge in Fig. \ref{fig:arc_map}). This global agreement seems to indicate that the shear is dominated by CGr32, and the small offset may be due to a nearby less massive sub-structure or to some asymmetry in CGr32 producing a local tilt of the external shear at the location of the lens. 
We also determine that the velocity dispersion of this massive structure is between 800~\kms\ and 1200~\kms , which is consistent with the velocity dispersion of CGr32 of $928$~\kms .

\section{Summary and conclusions}
\label{sec:conclusion}

We present the \textsc{Muse} gAlaxy Groups In \textsc{Cosmos} (MAGIC) survey, which was designed to investigate the impact of environment on galaxy evolution at intermediate redshift. It targets 14 known massive groups at $0.3<z<0.8$ in the COSMOS field, but its strength lies in the use of any galaxy in the 17 MUSE fields to span various environments, such as isolated galaxies, pairs, small groups, massive groups, and clusters.

Thanks to medium-deep exposures of 1-10 hours, and no a priori target selection, spectroscopic redshifts were computed for 1683 objects, including 1419 with secure redshifts, of which 1154 are galaxies in the redshift range of \oii\ emitters ($0.25\le z < 1.5$). The MAGIC survey increases the number of galaxies with known secure spectroscopic redshifts on the MAGIC fields  by a factor of five thanks to its ability to derive redshifts for objects with apparent magnitudes fainter than $z_{\rm app}^{++}>21.5$; previous spectroscopic campaigns missed these either because of selection or a lack of sensitivity.
The spectroscopic redshift completeness of the MAGIC sample is globally above 80\% in the redshift range $0.25\le z < 1.5$ down to an apparent magnitude of $z_{\rm app}^{++}=25.9$, and is locally close to unity down to $z_{\rm app}^{++}=23-24$ and above 50\% down to $z_{\rm app}^{++}=25.5$ in bins of redshift and apparent magnitude over the same redshift range.

With this paper, we release the MAGIC group and galaxy catalogs. Groups were identified through a FoF algorithm. Group Viral mass and radius were inferred from the velocity dispersion of the group members. For each galaxy, we determined the overdensity thanks to the VMC method on large-scale spectro-photometric surveys in COSMOS. For galaxies in groups, we further determined (i) another local density estimator using the Voronoi tessellations method based on MUSE spectroscopic redshifts at higher spatial resolution than the VMC method, and (ii) a global density using a phase-space parametrization based on the velocities of galaxies within groups and on their distance to the group center relative to group velocity dispersion and size.
The MAGIC group catalog contains 76 pairs of galaxies and 67 groups of at least three members, including 19 groups of ten or more members. The catalog stores all the global information about the groups. Most of the groups are in the redshift range $0.25\le z<1.5, $ where MUSE can detect the \oii\ doublet.
The galaxy catalog contains all 1683 objects with spectroscopic redshifts inferred from MUSE data. This catalog stores the redshift and position information as well as the identification of groups and the density estimators at the position of the galaxy. It further contains galaxy properties and rest-frame colors inferred from SED fitting with the \textsc{Cigale} software on the COSMOS photometry.

Owing to its observing strategy, MAGIC contains galaxies in dense, massive, and rich groups as well as galaxies in field and low-richness groups.
We showed that the fraction of quiescent galaxies is low ---namely of about $\sim 5$\%--- in regions of low overdensity, in groups of ten or less members, and in the outskirts of massive groups, where it is similar to that of the population of field galaxies.
This fraction significantly increases with local density and with group mass. For groups richer than ten members, the fraction is stable, but it clearly increases with the time spent by galaxies within the groups.
We also find that the mass distribution of quiescent galaxies is shifted towards systems more massive than $M_\star=10^{10}~M_\sun$, and that the fraction of both quiescent and star-forming galaxies more massive than $M_\star=10^{10}~M_\sun$ gets larger as density increases.
Moreover, the fact that the fraction of galaxies less massive than $M_\star=10^{9}~M_\sun$ decreases with environment density seems to indicate that those galaxies merged to form more massive ones as environments became denser.
Our results also point toward a clear morphological dichotomy between disk-dominated star-forming galaxies and bulge-dominated quiescent galaxies. The impact of environment on bulge and disk contributions of star-forming galaxies is not large and could be due to the larger stellar mass of such galaxies in dense environments.
These findings suggest that galaxies are preprocessed in groups of increasing mass before entering massive groups and clusters.

Several peculiar features found in groups were also highlighted. Two groups and one cluster at a similar redshift ($z\sim 0.73$) and with relatively small transverse separations are observed and may be connected, the two groups being either subgroups of the cluster or infalling groups along cosmic web filaments. We also found two massive groups along the same line of sight with a scaled redshift separation of $\Delta z = 0.01$, which is too large for the two groups to be connected. Surprisingly, only one of those groups was previously identified in the zCOSMOS 20k group survey.
The MAGIC dataset also enabled us to serendipitously find new ionized gas nebulae within massive groups, probably associated to interaction between galaxies or between galaxies and the intragroup medium. In addition, one of the MAGIC fields contains a quasar that is surrounded by an extended ionized gas nebula that resembles other quasar-illuminated nebulae, except that the quasar in MAGIC is about ten times fainter, with a SMBH that is ten times less massive than other such quasars. Finally, a gravitational arc is also present in one MAGIC field and the MUSE data allowed us to find four images of a \lya\ emitter at $z\sim 4$ lensed by a massive galaxy at $z\sim 0.725$.

All these findings show that the large monolithic FoV of the MUSE spectrograph is well suited to identifying associations between galaxies down to low mass and large-scale structures, and to studying the interconnection between galaxies and their environment.
In the evolving context of galaxy evolution with environment, integral field spectrographs with larger FoVs would be necessary to map the cosmic web filaments in order to better explore the connections between groups.

\begin{acknowledgements}
We dedicate this article in memory of Hayley Finley.
This work has been carried out through the support of the ANR FOGHAR (ANR-13-BS05-0010-02), the OCEVU Labex (ANR-11-LABX-0060), and the A*MIDEX project (ANR-11-IDEX-0001-02), which are funded by the ``Investissements d'avenir'' French government program managed by the ANR.
BCL acknowledges support from the National Science Foundation under Grant No. 1908422.
JB acknowledges support by Funda{\c c}{\~a}o para a Ci{\^e}ncia e a Tecnologia (FCT) through national funds (UID/FIS/04434/2013) and Investigador FCT contract IF/01654/2014/CP1215/CT0003., and by FEDER through COMPETE2020 (POCI-01-0145-FEDER-007672).
NFB acknowledges support from the ANR grant `3DGasFlows' (ANR-17-CE31-0017).
VAM acknowledges the Ministry of Science, Technology and Innovation of Colombia (MINCIENCIAS) PhD fellowship program No. 756-2016.
TU acknowledges funding from the ERC Advanced Grant SPECMAP-CGM-101020943.
\end{acknowledgements}

\bibliographystyle{aa}
\bibliography{survey.bib}

\appendix
\section{Catalogs and data release}
\label{app:catalogs}

With this publication, we publicly release two catalogs for groups and galaxies, respectively, in the MAGIC sample.
We also make the 18 fully reduced datacubes\footnote{This includes the VVDS group that is not used in the present analysis.} available on demand without any condition, by contacting the first or second author of this paper. They are also stored on the ESO archive\footnote{\url{http://archive.eso.org}}, together with the corresponding MUSE raw data.

The MAGIC group catalog has 143 entries, including 76 pairs and 67 groups. It summarizes the main properties of groups inferred mostly in Sect. \ref{sec:group_properties}. The description of the columns included this catalog is given in Table \ref{tab:group_catalog}.

The MAGIC galaxy catalog contains 2471 entries, corresponding to all objects having a tentative redshift within MUSE data, but also to objects in the COSMOS2015 photometric parent sample \citep{Laigle+16} with no redshift inferred from MUSE data. Table \ref{tab:group_catalog} describes the columns of this catalog, including proper references to the sections where the corresponding properties are explained.

\begin{table*}
\caption{\label{tab:group_catalog}Description of the columns of the MAGIC group catalog. The properties are described in Sect. \ref{sec:group_properties}.}
\footnotesize
\centering
 \begin{tabular}{ll}
 \hline
 \hline
  Col. name & Description\\
 \hline
  MGR\_ID & MAGIC group ID \\
  RA & Right ascension (J2000, degrees) of the unweighted barycenter of the group\\
  DEC & Declination (J2000, degrees) of the unweighted barycenter of the group\\
  RICH & Number of members within the group \\
  Z & Mean redshift of the group\\
  SIGMA & Velocity dispersion $\sigma_g$ (\kms ) of galaxies in the group, using the Gapper method\\
  LMVIR & Virial mass $M_{\rm Vir}$ in log ($M_\sun$)\\
  R200 & $R_{200}$ radius (kpc)\\
 \hline
  CGR\_ID & Group ID in the zCOSMOS 20k group catalog \citep{Knobel+12}, when matched\\
  CW\_ID & Group ID in the COSMOS Wall group catalog \citep{Iovino+16}, when matched\\
  XRAY\_ID & Group ID in the X-ray group catalog of \cite{Gozaliasl+19}, when matched\\
  XRAY\_SEP & Angular separation (arcsecond) between MAGIC and X-ray \citep{Gozaliasl+19} centers\\
  XLM200 & Mass at radius $R_{200}$ in log ($M_\sun$) inferred from X-rays \citep{Gozaliasl+19} \\
  XR200 & $R_{200}$ radius (kpc) inferred from X-rays \citep{Gozaliasl+19} \\
  \hline
 \end{tabular}
\tablefoot{Velocity dispersion $\sigma_g$, radius $R_{200}$, and Virial mass $M_{\rm Vir}$ are only provided for groups of more than five members.}
\end{table*}

\begin{table*}
\caption{\label{tab:galaxy_catalog}Description of the columns of the MAGIC galaxy catalog.}
\footnotesize
\centering
 \begin{tabular}{ll}
 \hline
 \hline
  Col. name & Description\\
 \hline
  ID & MAGIC object ID, of the form X\_CGrY, where X is the identifier within the fields corresponding to COSMOS group CGrY \\
  C2015\_ID & Object ID in the COSMOS2015 catalog \citep{Laigle+16} \\
  FIELD & Name of the MAGIC field, see Table \ref{tab:log}\\
  RA & Right ascension (J2000, degrees), see Sect. \ref{sec:data-reduction}, adjusted with the morphology analysis \citep{Mercier+23}, when available \\
  DEC & Declination in (J2000, degrees), see Sect. \ref{sec:data-reduction}, adjusted with the morphology analysis \citep{Mercier+23}, when available \\
  Z & Systemic redshift of the source, adjusted with the kinematics analysis \citep{Mercier+23}, when available, see Sect. \ref{sec:zmeasurements} \\
  CONFID & Redshift confidence level from 1 (low) to 3 (high), see Sect. \ref{sec:zmeasurements}\\
 \hline
  LMS & Logarithm of stellar mass $M_\star$ (in $M_\sun$) within 3\arcsec\ diameter apertures estimated from SED fitting with CIGALE, see Sect. \ref{sec:sed_fitting} \\
  DLMS & Uncertainty on the logarithm of stellar mass (in $M_\sun$) within 3\arcsec\ diameter apertures \\
  LSFR & Logarithm of SFR (in $M_\sun~\textrm{yr}^{-1}$) within 3\arcsec\ diameter apertures estimated from SED fitting with CIGALE, see Sect. \ref{sec:sed_fitting} \\
  DLSFR & Uncertainty on the logarithm of SFR (in $M_\sun~\textrm{yr}^{-1}$) within 3\arcsec\ diameter apertures\\
  E(B-V) & Color excess estimated from SED fitting with CIGALE, see Sect. \ref{sec:sed_fitting}\\
  NUV-RP & $NUV-r^{+}$ rest-frame color estimated from SED fitting with CIGALE, see Sect. \ref{sec:sed_fitting}\\
  RP-J & $r^{+}-J$ rest-frame color estimated from SED fitting with CIGALE, see Sect. \ref{sec:sed_fitting}\\
 \hline
  MGR\_ID & MAGIC group ID to which the galaxy belongs, as defined in the group catalog, see Table \ref{tab:group_catalog} \\
  VMC\_D & Galaxy density $\Sigma$ in log (Mpc$^{-2}$) at the location of the galaxy inferred from the VMC method, see Sect. \ref{sec:vmc}\\
  VMC\_OD & Galaxy over-density ($1+\delta$) in log at the location of the galaxy inferred from the VMC method, see Sect. \ref{sec:vmc}\\
  VOR\_D & Galaxy density in log (Mpc$^{-2}$) at the location of the galaxy inferred from Voronoi tessellations, see Sect. \ref{sec:muse_voronoi}\\
  DR & Distance to group center $\Delta r$ (kpc), see Sect. \ref{sec:eta}\\
  DV & Velocity $\Delta v$ (\kms ) in the frame of the group with respect to the group redshift, see Sect. \ref{sec:eta}\\
  ETA & Phase space parameter $\eta$, see Sect. \ref{sec:eta}\\
 \hline
 \end{tabular}
 \tablefoot{Stars are included in the catalog, but their physical and environment properties are not computed. Such properties are not provided either for the 788 galaxies without MUSE redshift determination.}
\end{table*}

\end{document}